\documentclass[twocolumn]{aastex631}
\usepackage{multirow}
\usepackage{subfigure}
\usepackage{booktabs}
\usepackage{graphicx}
\usepackage{amsmath}
\usepackage{CJK}
\usepackage{xcolor} 

\hypersetup{hypertex=true,
	colorlinks=true,
	linkcolor=blue,
	anchorcolor=blue,
	citecolor=blue}

\shorttitle{Chemical clock of N$_{2}$H$^{+}$/CCS}
\shortauthors{Chen et al. 2024}

\graphicspath{{./}{figures/}}

\begin{document}
\title{The Chemical Clock of High-mass Star-forming Regions: N$_{2}$H$^{+}$/CCS}

\correspondingauthor{JiangShui Zhang}
\email{jszhang@gzhu.edu.cn}

\author[0000-0001-8980-9663]{J. L. Chen}
\affil{Center for Astrophysics, Guangzhou University, Guangzhou, 510006, PR China} 

\author[0000-0002-5161-8180]{J. S. Zhang}
\affil{Center for Astrophysics, Guangzhou University, Guangzhou, 510006, PR China} 

\author[0000-0002-9600-1846]{J. X. Ge}
\affil{Xinjiang Astronomical Observatory, Chinese Academy of Sciences, Urumqi 830011, China} 
\affil{Xinjiang Key Laboratory of Radio Astrophysics, 150 Science1-Street, Urumqi 830011, China} 

%


\author[0000-0001-9155-0777]{Y. X. Wang}
\affil{Center for Astrophysics, Guangzhou University, Guangzhou, 510006, PR China} 

\author[0000-0002-5634-131X]{H. Z. Yu}
\affil{Ural Federal University, 19 Mira Street, 620002 Ekaterinburg, Russia} 

\author[0000-0002-5230-8010]{Y. P. Zou}
\affil{Center for Astrophysics, Guangzhou University, Guangzhou, 510006, PR China} 

%
%
%
\author[0000-0001-5574-0549]{Y. T. Yan}
\affil{Max-Planck-Institut f{\"u}r Radioastronomie, Auf dem H{\"u}gel 69, D-53121 Bonn, Germany}

\author[0009-0008-8866-0287]{X. Y. Wang}
\affil{Center for Astrophysics, Guangzhou University, Guangzhou, 510006, PR China} 

\author[0009-0003-9805-7901]{D. Y. Wei}
\affil{Center for Astrophysics, Guangzhou University, Guangzhou, 510006, PR China} 


\begin{abstract}

Using the IRAM 30 m telescope, we presented observations of N$_{2}$H$^{+}$ $J$ = 1$-$0, CCS $J_{N}$ = $8_{7}-7_{6}$ and $7_{7}-6_{6}$ lines toward a large sample of ultracompact H{\footnotesize II} regions (UC H{\footnotesize II}s). Among our 88 UC H{\footnotesize II}s, 87 and 33 sources were detected in the N$_{2}$H$^{+}$ $J$ = 1$-$0 and CCS $J_{N}$ = $8_{7}-7_{6}$ lines, respectively. For the CCS $7_{7}-6_{6}$ transition, we detected emission in 10 out of 82 targeted sources, all of which also exhibited emission in the CCS $J_{N}$ = $8_{7}-7_{6}$ line. Physical parameters are derived for our detections, including the optical depth and excitation temperature of N$_{2}$H$^{+}$, the rotational temperature of CCS and the column density. Combining our results and previous observation results in different stages of high-mass star-forming regions (HMSFRs), we found that the column density ratio $N$(N${_2}$H$^{+}$)/$N$(CCS) increases from high-mass starless cores (HMSCs) through high-mass protostellar cores (HMPOs) to UC H{\footnotesize II}s. This implies that $N$(N${_2}$H$^{+}$)/$N$(CCS) can trace the evolution process of HMSFRs. It was supported by our gas-grain chemical model, which shows that $N$(N${_2}$H$^{+}$)/$N$(CCS) increases with the evolution age of HMSFRs. 
The temperature, density and chemical age were also constrained from our best-fit model at each stage. 
Thus, we propose $N$(N${_2}$H$^{+}$)/$N$(CCS) as a reliable chemical clock of HMSFRs.

\end{abstract}


\section{Introduction} \label{sec:Introduction}

High-mass stars (\textgreater 8 $M_\odot$ and \textgreater 10$^{3}$ $L_\odot$) dominate the energy budget of galaxies, influencing their evolution and, ultimately, that of the Universe as a whole \citep{1996GReGr..28.1309M,2012ARA&A..50..531K}. They produce heavy elements, which are fed back into the interstellar medium through their strong stellar winds and supernovae \citep[e.g.,][]{2007ARA&A..45..565M}. This process alters the composition and chemistry of their local environments, providing the raw material for subsequent generations of star and planet formation \citep[e.g.,][]{2007ARA&A..45..481Z,2014prpl.conf..243K}.  Despite their pivotal role, the formation of high-mass stars remains an enigmatic process, being far less understood \citep[e.g.,][]{2018ARA&A..56...41M,2023MNRAS.524.4384P}. 
Therefore, it is important to understand thoroughly the high-mass star formation process.


Currently, there is widespread consensus that high-mass star-forming regions (HMSFRs) evolve from high-mass starless cores (HMSCs) to high-mass protostellar cores (HMPOs) and then to ultracompact H{\footnotesize II} regions (UC H{\footnotesize II}s) \citep[e.g.,][]{2007prpl.conf..165B,2015A&A...576A.131H,2023ApJS..264...48W}. HMSCs, the initial phase of HMSFRs, originate from cold, dense gas and dust at $\sim$10 K, emitting primarily at submillimeter wavelengths with minimal mid-infrared emission \citep[e.g.,][]{2005ApJ...634L..57S,2006ApJ...641..389R,2010A&A...518L..78B,2017ApJS..231...11Y}. HMPOs represent the second phase of HMSFRs, marking the formation of active protostar(s), which emit emission at mid-infrared wavelengths, but no radio continuum emission. During this phase, a formed disk transfers the infalling material from the envelope to the HMPO \cite[e.g.,][]{2002ApJ...566..931S,2002ApJ...566..945B,2006ApJ...638..241E,2007prpl.conf..197C,2009A&A...498..147G,2021ApJ...912..108H}. As HMPOs continue to evolve, the mass and luminosity of HMPOs increase, resulting in an elevated surface temperature that generates vast amounts of Lyman continuum photons. These high energy photons ionize the surrounding material, leading to the formation of bubbles of hot ionized gas, known as UC H{\footnotesize II}s. These regions experience higher pressure compared to the surrounding gas, resulting in rapid expansion, which propel ionization shocks into the surrounding medium and compress the adjacent material \citep{1989ApJS...69..831W,1995MNRAS.277..700D,2002ARA&A..40...27C}. This process possibly triggers the emergence of a new generation of stars \citep{1994MNRAS.268..291W,2012MNRAS.421..408T}.
However, such classification is rather simplistic and may result in overlapping classifications among these stages \citep[e.g.,][]{2018ARA&A..56...41M}.

For comprehensively understanding the HMSFRs, determining its chemical composition at different evolution stages is necessary, since chemical composition contains various information in it \citep[e.g.,][]{2010PASJ...62.1473T,2011ARA&A..49..471M,2012A&ARv..20...56C,2018IAUS..332....3V,2014PASJ...66..119O,2020ARA&A..58..727J,2021PhR...893....1O,2022MNRAS.510.3389U}. The ratio of column densities of two molecules can be used to investigate the chemical evolution of HMSFRs, which is known as the chemical clock of HMSFRs \citep[e.g.,][]{2004A&A...422..159W,2012ApJ...756...60S,2017ApJS..228...12T,2019ApJ...872..154T,2019A&A...622A..32L,2021SCPMA..6479511X,2023ApJS..264...48W}.
Using measurements on proposed different tracers, many works focused on investigating individual stage of HMSFRs, such as HMSCs \citep{2010ApJ...714.1658S,2011A&A...527A..88V,2012ApJ...751..105V,2012ApJ...756...60S,2013ApJ...773..123S}, HMPOs \citep{2002ApJ...566..945B,2018ApJS..236...45G} and UC H{\footnotesize II} \citep{1998A&AS..133...29H,2007A&A...465..219P}. In addition, some works tried to investigate chemical clock for tracing evolution of HMSFRs. Based on observations of N$_{2}$H$^{+}$ and HCO$^{+}$ toward HMSFRs, \cite{2013ApJ...777..157H} found that the column density ratio of N$_{2}$H$^{+}$/HCO$^{+}$ rises modestly from HMSCs to HMPOs, and further to UC H{\footnotesize II}s. \cite{2015MNRAS.451.2507Y} investigated the ratios of $N$(N$_{2}$H$^{+}$)/$N$(H$^{13}$CO$^{+}$) and $N$(CCH)/$N$(H$^{13}$CO$^{+}$), revealing a marginal decrease from HMPOs to UC H{\footnotesize II}s. Based on H$^{13}$CN and HN$^{13}$C line measurements, \cite{2015ApJS..219....2J} identified a slight statistically increasing trend in the HCN/HNC abundance ratio with the evolution of HMSFRs. \cite{2019ApJ...872..154T} carried out one survey on HC$_{3}$N ($J$ = 9 -- 8 and 10 -- 9), N$_{2}$H$^{+}$ ($J$ = 1 -- 0), and CCS ($J_{N}$ = 7$_{6}$ -- 6$_{5}$) toward HMSCs and HMPOs, showing that the $N$(HC$_{3}$N)/$N$(N$_{2}$H$^{+}$) ratio increases from HMSCs to HMPOs. Further,  \cite{2023ApJS..264...48W} found an increasing trend of $N$(HC$_{3}$N)/$N$(N$_{2}$H$^{+}$) from HMSCs, HMPOs to UC H{\footnotesize II}s, based on observations toward UC H{\footnotesize II}s and data in HMSC and HMPO from \cite{2019ApJ...872..154T}. Using the molecular lines of N$_{2}$H$^{+}$ and CCS, the relative low values of $N$(N$_{2}$H$^{+}$)/$N$(CCS) were obtained in HMSCs \citep{2011AA...529L...7F,2023AA...680A..58F}, while much higher values of the ratio were reported in HMPOs \citep{2019ApJ...872..154T}. The significant difference in the ratio of N$_{2}$H$^{+}$/CCS in the HMSCs and HMPOs stages suggests that N$_{2}$H$^{+}$/CCS could serve as a potential chemical clock in HMSFRs. To check if the $N$(N$_{2}$H$^{+}$)/$N$(CCS) ratio can be taken as a good chemical clock, we conducted observations toward HMSFRs at another evolution stage, i.e., UC H{\footnotesize II}s, targeting the transition lines of N$_{2}$H$^{+}$ $J$ = 1$-$0, CCS $J_{N}$ = $8_{7}-7_{6}$ and $7_{7}-6_{6}$.

In this paper, we presented observations on N$_{2}$H$^{+}$ and CCS toward a sample of UC H{\footnotesize II}s using the Institut de Radioastronomie Millmétrique (IRAM) 30 m telescope\footnote{\url{https://iram-institute.org/observatories/30-meter-telescope/}}. The details of our sample and observations are summarized in Section \ref{sec:Sample Selection and observations}. In Section \ref{sec:Results}, the fitting results for the detected spectral lines are presented, along with estimation for the physical parameters of our sample, including the optical depth, column density and temperature. Section \ref{sec:Discussion} explores the $N$(N${2}$H$^{+}$)/$N$(CCS) ratio across distinct stages of HMSFRs and models the chemical evolution with a gas-grain chemical model. The goal is to assess the viability of the $N$(N$_{2}$H$^{+}$)/$N$(CCS) ratio as a reliable chemical clock in the context of HMSFRs. A concise summary is provided in Section \ref{sec:summary}.

\section{Sample Selection and observations} \label{sec:Sample Selection and observations}

\subsection{Sample Selection}\label{sec:Sample distance}


UC H{\footnotesize II}s are the most luminous objects in the Milky Way at far-infrared wavelengths accompanied radio continuum emission. Based on an infrared survey \citep{2001AJ....121.2819P}, radio surveys \citep{2012PASP..124..939H,2013ApJS..205....1P}, and submillimeter surveys \citep{2006A&A...453.1003T,2009A&A...504..415S,2014A&A...565A..75C}, over 1000 UC H{\footnotesize II}s has been confirmed \citep[e.g.,][]{2005AJ....129..348G,2005AJ....130..156G,2007A&A...461...11U,2009AA501539U,2013MNRAS.435..400U,2014AA567L5D,2015A&A...579A..71C,2016ApJ...833...18H,2018A&A...615A.103K,2019MNRAS.487.1057D,2019ApJS..244...35L,2023MNRAS.520.1073I}. 

Thanks to the Bar and Spiral Structure Legacy (BeSSeL\footnote{\url{http://bessel.vlbi-astrometry.org/}}) project \citep{2014ApJ...783..130R,2019ApJ...885..131R}, a large sample of 199 HMSFRs with maser emission have been measured with accurate distance values. We have performed systematic observation  survey on C$^{18}$O (C$^{17}$O), NH$_3$ ($^{15}$NH$_3$), CN ($^{15}$CN) and CS (C$^{34}$S, $^{13}$CS, C$^{33}$S) toward this sample, to measure the Galactic  interstellar carbon, nitrogen, oxygen and sulphur isotope ratios \citep{2015ApJS..219...28Z,2016RAA....16...47L,2019ApJ...877..154Y,2020ApJ...899..145Y,2020ApJS..249....6Z,2023ApJS..268...56Z,2021ApJS..257...39C,2024ApJ...971..164C}. Crossmatching this HMSFR sample with those known UC H{\footnotesize II}s, we obtain 152 UC H{\footnotesize II}s with accurate distance values. Among these 152 UC H{\footnotesize II}s, 88 sources with strong CS emission \citep[the main beam brightness peak temperature \textgreater 0.6K,][]{2020ApJ...899..145Y} were selected as our sample (Table \ref{tab:source list}), which should improve the detection rate of CCS.


\startlongtable
\renewcommand\tabcolsep{25.0pt} 
		\begin{deluxetable*}{lcccc}
			\tablecaption{Detailed Information of Our UC H{\footnotesize II} Sample \label{tab:source list}}
			\tablehead{
				\colhead{Object}&
				\colhead{$\alpha$(2000)}&
				\colhead{$\delta$(2000)}&
				\colhead{D$_{sun}$}&
				\colhead{References}
				\\
				\colhead{}&
				\colhead{($^h \; ^m \; ^s$)}&
				\colhead{($^{\circ} \; ^{\prime} \; ^{\prime\prime}$)}&
				\colhead{(kpc)}&
				\colhead{}
			}
			\decimalcolnumbers
			\startdata
			G133.94+01.06 & 02:27:04.18 & 61:52:25.4  & 1.95  $\pm$ 0.04  & {[}1{]}  \\
			G359.13       & 17:43:26.00 & -29:39:17.5 & 6.06  $\pm$ 1.14  & {[}2{]}  \\
			G359.61       & 17:45:39.07 & -29:23:30.2 & 2.67  $\pm$ 0.15  & {[}3{]}  \\
			G000.31       & 17:47:09.11 & -28:46:16.2 & 2.92  $\pm$ 0.36  & {[}2{]}  \\
			G001.1-00.1   & 17:48:42.24 & -28:01:27.7 & 7.76  $\pm$ 0.07  & {[}2{]}  \\
			G001.14       & 17:48:48.54 & -28:01:11.3 & 5.15  $\pm$ 4.28  & {[}2{]}  \\
			G001.00       & 17:48:55.29 & -28:11:48.2 & 11.11  $\pm$ 7.04 & {[}2{]}  \\
			G002.70       & 17:51:45.98 & -26:35:57.0 & 9.90  $\pm$ 10.29 & {[}2{]}  \\
			G006.79       & 18:01:57.75 & -23:12:34.2 & 3.47  $\pm$ 0.25  & {[}2{]}  \\
			G007.47       & 18:02:13.18 & -22:27:58.9 & 20.41  $\pm$ 2.50 & {[}4{]}  \\
			G009.62+00.19 & 18:06:14.13 & -20:31:44.3 & 5.15  $\pm$ 0.61  & {[}5{]}  \\
			G009.21       & 18:06:52.84 & -21:04:27.8 & 3.30  $\pm$ 1.05  & {[}2{]}  \\
			G010.32       & 18:09:01.46 & -20:05:07.8 & 3.53  $\pm$ 0.72  & {[}2{]}  \\
			G010.62       & 18:10:17.99 & -19:54:04.6 & 4.95  $\pm$ 0.47  & {[}2{]}  \\
			G011.10       & 18:10:28.25 & -19:22:30.2 & 2.75  $\pm$ 0.20  & {[}2{]}  \\
			G012.81-00.19 & 18:14:14.39 & -17:55:49.9 & 2.92  $\pm$ 0.31  & {[}6{]}  \\
			G013.71       & 18:15:36.98 & -17:04:32.1 & 3.79  $\pm$ 0.20  & {[}2{]}  \\
			G018.34       & 18:17:58.13 & -12:07:24.8 & 2.00  $\pm$ 0.08  & {[}2{]}  \\
			G015.66       & 18:20:59.75 & -15:33:09.8 & 4.55  $\pm$ 0.60  & {[}2{]}  \\
			G017.63       & 18:22:26.38 & -13:30:11.9 & 1.49  $\pm$ 0.04  & {[}2{]}  \\
			G019.00       & 18:25:44.78 & -12:22:45.8 & 4.05  $\pm$ 1.03  & {[}2{]}  \\
			G019.49       & 18:26:09.17 & -11:52:51.3 & 3.07  $\pm$ 0.94  & {[}2{]}  \\
			G019.36       & 18:26:25.78 & -12:03:53.2 & 2.92  $\pm$ 0.59  & {[}2{]}  \\
			G016.86       & 18:29:24.41 & -15:16:04.1 & 2.35  $\pm$ 0.51  & {[}2{]}  \\
			G017.02       & 18:30:36.29 & -15:14:28.3 & 1.88  $\pm$ 0.38  & {[}2{]}  \\
			G022.35       & 18:31:44.12 & -09:22:12.3 & 4.33  $\pm$ 2.02  & {[}2{]}  \\
			G023.38       & 18:33:14.32 & -08:23:57.5 & 4.81  $\pm$ 0.58  & {[}2{]}  \\
			G023.25       & 18:34:31.24 & -08:42:47.3 & 5.92  $\pm$ 1.79  & {[}2{]}  \\
			G023.43       & 18:34:39.19 & -08:31:25.4 & 5.88  $\pm$ 1.11  & {[}2{]}  \\
			G023.20       & 18:34:55.18 & -08:49:15.2 & 4.18  $\pm$ 0.60  & {[}2{]}  \\
			G024.78       & 18:36:12.56 & -07:12:10.8 & 6.67  $\pm$ 0.71  & {[}2{]}  \\
			G024.85       & 18:36:18.39 & -07:08:50.8 & 5.68  $\pm$ 0.52  & {[}2{]}  \\
			G024.63       & 18:37:22.71 & -07:31:42.0 & 4.13  $\pm$ 0.77  & {[}2{]}  \\
			G028.14       & 18:42:42.59 & -04:15:35.1 & 6.33  $\pm$ 0.92  & {[}2{]}  \\
			G028.39       & 18:42:51.98 & -03:59:54.4 & 4.33  $\pm$ 0.28  & {[}2{]}  \\
			G028.30       & 18:44:21.97 & -04:17:39.9 & 4.52  $\pm$ 0.45  & {[}2{]}  \\
			G028.83       & 18:44:51.09 & -03:45:48.3 & 5.00  $\pm$ 1.00  & {[}2{]}  \\
			G030.78       & 18:46:48.09 & -01:48:53.9 & 7.14  $\pm$ 1.63  & {[}2{]}  \\
			G030.19       & 18:47:03.07 & -02:30:36.2 & 4.72  $\pm$ 0.22  & {[}2{]}  \\
			G030.22       & 18:47:08.30 & -02:29:29.3 & 3.52  $\pm$ 0.40  & {[}2{]}  \\
			G030.70       & 18:47:36.80 & -02:00:54.3 & 6.54  $\pm$ 0.85  & {[}2{]}  \\
			G030.74       & 18:47:39.73 & -01:57:24.9 & 3.07  $\pm$ 0.52  & {[}2{]}  \\
			G030.41       & 18:47:40.76 & -02:20:30.9 & 3.95  $\pm$ 0.33  & {[}2{]}  \\
			G030.81       & 18:47:46.98 & -01:54:26.4 & 3.12  $\pm$ 0.36  & {[}2{]}  \\
			G031          & 18:48:12.39 & -01:26:30.7 & 5.43  $\pm$ 0.50  & {[}2{]}  \\
			G030.97       & 18:48:22.04 & -01:48:30.7 & 3.40  $\pm$ 0.25  & {[}2{]}  \\
			G031.24       & 18:48:45.08 & -01:33:13.2 & 13.16  $\pm$ 2.42 & {[}7{]}  \\
			G032.79       & 18:50:30.73 & -00:01:59.2 & 9.71  $\pm$ 2.92  & {[}7{]}  \\
			G032.74       & 18:51:21.86 & -00:12:06.2 & 7.94  $\pm$ 1.01  & {[}2{]}  \\
			G033.39       & 18:52:14.64 & 00:24:54.3  & 8.85  $\pm$ 2.27  & {[}2{]}  \\
			G034.41       & 18:53:18.03 & 01:25:25.5  & 2.94  $\pm$ 0.10  & {[}2{]}  \\
			G34.3+0.2     & 18:53:18.40 & 01:14:56.0  & 9.75  $\pm$ 0.37  & {[}2{]}  \\
			G033          & 18:53:32.56 & 00:31:39.1  & 7.63  $\pm$ 1.17  & {[}2{]}  \\
			G037.42       & 18:54:14.35 & 04:41:39.6  & 1.88  $\pm$ 0.07  & {[}2{]}  \\
			G036.11       & 18:55:16.79 & 03:05:05.3  & 4.07  $\pm$ 0.93  & {[}2{]}  \\
			G035.79       & 18:57:16.89 & 02:27:58.0  & 8.85  $\pm$ 1.02  & {[}2{]}  \\
			G035.14       & 18:58:12.62 & 01:40:50.5  & 2.19  $\pm$ 0.22  & {[}2{]}  \\
			G034.79       & 18:59:45.98 & 01:01:18.9  & 2.62  $\pm$ 0.14  & {[}2{]}  \\
			G037.47       & 19:00:07.14 & 03:59:52.9  & 11.36  $\pm$ 3.87 & {[}2{]}  \\
			G038.11       & 19:01:44.15 & 04:30:37.4  & 4.13  $\pm$ 0.60  & {[}2{]}  \\
			G038.03       & 19:01:50.47 & 04:24:18.9  & 10.53  $\pm$ 2.44 & {[}2{]}  \\
			G040.42       & 19:02:39.62 & 06:59:09.0  & 12.82  $\pm$ 2.14 & {[}2{]}  \\
			G040.28       & 19:05:41.22 & 06:26:12.7  & 3.37  $\pm$ 0.22  & {[}2{]}  \\
			G040.62       & 19:06:01.63 & 06:46:36.1  & 12.50  $\pm$ 3.28 & {[}2{]}  \\
			G041.22       & 19:07:21.38 & 07:17:08.1  & 8.85  $\pm$ 1.72  & {[}2{]}  \\
			G042.03       & 19:07:28.18 & 08:10:53.4  & 14.08  $\pm$ 2.38 & {[}2{]}  \\
			G043.03       & 19:11:38.98 & 08:46:30.6  & 7.69  $\pm$ 1.12  & {[}2{]}  \\
			G045.49       & 19:14:11.36 & 11:13:06.3  & 6.94  $\pm$ 1.16  & {[}2{]}  \\
			G045.45       & 19:14:21.27 & 11:09:15.8  & 8.40  $\pm$ 1.20  & {[}7{]}  \\
			G043          & 19:14:26.39 & 09:22:36.5  & 7.46  $\pm$ 0.72  & {[}8{]}  \\
			G045.80       & 19:16:31.08 & 11:16:11.9  & 7.30  $\pm$ 1.23  & {[}2{]}  \\
			G049.34       & 19:20:32.45 & 14:45:45.3  & 4.15  $\pm$ 0.53  & {[}2{]}  \\
			G049.26       & 19:20:44.86 & 14:38:26.8  & 8.85  $\pm$ 1.25  & {[}2{]}  \\
			G049.41       & 19:20:59.21 & 14:46:49.6  & 7.58  $\pm$ 1.78  & {[}2{]}  \\
			G048.99       & 19:22:26.14 & 14:06:39.1  & 5.62  $\pm$ 0.54  & {[}7{]}  \\
			G049.59       & 19:23:26.61 & 14:40:16.9  & 4.59  $\pm$ 0.19  & {[}2{]}  \\
			G049.04       & 19:25:22.25 & 13:47:19.5  & 6.10  $\pm$ 0.82  & {[}2{]}  \\
			G058.77       & 19:38:49.13 & 23:08:40.2  & 3.34  $\pm$ 0.45  & {[}2{]}  \\
			G059.83       & 19:40:59.29 & 24:04:44.1  & 4.13  $\pm$ 0.24  & {[}2{]}  \\
			G059          & 19:43:11.25 & 23:44:03.3  & 2.16  $\pm$ 0.09  & {[}9{]}  \\
			G060.57       & 19:45:52.50 & 24:17:43.2  & 8.26  $\pm$ 1.02  & {[}2{]}  \\
			G070.18       & 20:00:54.49 & 33:31:28.2  & 6.41  $\pm$ 0.66  & {[}2{]}  \\
			G071.52       & 20:12:57.89 & 33:30:27.0  & 3.61  $\pm$ 0.34  & {[}2{]}  \\
			G090.92       & 21:09:12.97 & 50:01:03.6  & 5.85  $\pm$ 1.06  & {[}2{]}  \\
			G097.53       & 21:32:12.43 & 55:53:49.6  & 7.52  $\pm$ 0.96  & {[}2{]}  \\
			G108.18+05.51 & 22:28:52.20 & 64:13:43.0  & 0.78  $\pm$ 0.09  & {[}2{]}  \\
			G108.20       & 22:49:31.47 & 59:55:42.0  & 4.41  $\pm$ 0.72  & {[}7{]}  \\
			G109.87       & 22:56:18.00 & 62:01:49.5  & 0.70  $\pm$ 0.04  & {[}2{]} \\
			\enddata
			\tablecomments{
				Column(1): source name; column(2): R.A. (J2000); column(3): decl. (J2000); column(4): the heliocentric distance; 
				column(5): references. 
				[1]\cite{2014AA567L5D}, [2] \cite{2016ApJ...833...18H}, [3] \cite{2019ApJS..244...35L}, [4] \cite{2005AJ....129..348G}, [5] \cite{2004AA417615C}, [6] \cite{2022AA664A140K}, [7] \cite{2009AA501539U}, [8] \cite{2001AA3761064V}, [9] \cite{2022ApJS..258...19S}.
				}
		\end{deluxetable*}

\subsection{Observations}\label{sec:Observations}

The observations of  the  N$_{2}$H$^{+}$ $J$ = 1$-$0 and CCS $J_{N}$ = $8_{7}-7_{6}$ lines were observed simultaneously toward 10 UC H{\footnotesize II}s (including 4 sources being observed in CCS $7_{7}-6_{6}$ line) in 2016 June within project 013-16, with the Institut de Radio Astronomie Millimétrique (IRAM) 30 m single dish telescope\footnote{The IRAM 30 m is supported by Institut National des Sciences de L’univers/Centre National de la Recherche Scientifique, (INSU/CNRS, France), Max-Planck-Gesellschaft (MPG, Germany), and Instituto Geográfico Nacional (IGN, Spain).}, at the Pico Veleta Observatory (Granada, Spain). In 2020 August, we performed observations toward the other 78 sources within project 022-20 with IRAM 30 m telescope. During this observations, the N$_{2}$H$^{+}$ $J$ = 1$-$0, CCS $J_{N}$ = $8_{7}-7_{6}$ and CCS $7_{7}-6_{6}$ lines were observed simultaneously. In summary, 88 UC H{\footnotesize II}s were observed in N$_{2}$H$^{+}$ $J$ = 1$-$0 and CCS $J_{N}$ = $8_{7}-7_{6}$ lines, while 82 sources were observed in CCS $7_{7}-6_{6}$ lines. 
The rest frequencies of N$_{2}$H$^{+}$ $J$ = 1$-$0, CCS $J_{N}$ = $8_{7}-7_{6}$ and  $7_{7}-6_{6}$ lines are 93173.700, 93870.107 and 90686.381 MHz, respectively, with a corresponding beam size of $\sim$27\arcsec (Table \ref{tab:linePara}). 

The Eight Mixer Receiver (EMIR) with dual-polarization and the Fourier Transform Spectrometers (FTS) backend were used, providing a spectral resolution of $\sim$0.6 km s$^{-1}$ around 93 GHz. The standard position switching mode was carried out with the off position at a (-30$^\prime$, 0$^\prime$) or (30$^\prime$, 0$^\prime$) offset in R.A. and decl. from the source. The on-source integration time depended on the line intensity, with an integration time ranging from 4 to 120 minutes (Table \ref{tab:obserUCHII}). We checked the pointing every two hours toward nearby strong continuum sources (e.g., 3C 123, or NGC 7027). Focus calibrations were done at the beginning of the observations and during sunset and sunrise toward strong quasars \citep{2023A&A...670A..98Y}. The system temperatures were 200-300 K on a antenna temperature ($T_{\rm A}^{*}$) scale for the observations, with an {\it rms} noise of 10-110 mK (Table \ref{tab:obserUCHII}). The main beam brightness temperature ($T_{\rm mb}$) was obtained from the $T_{\rm A}^{*}$ by multipling the ratio of the forward and main beam efficiencies (Feff/Beff $\sim$ 0.94/0.78 = 1.21) \footnote{\url{https://publicwiki.iram.es/Iram30mEfficiencies}}.

\renewcommand\tabcolsep{14.0pt} 
\begin{deluxetable*}{lcccccc}[htbp]
	\tablecaption{Theoretical line Parameters for N$_{2}$H$^{+}$ ($J$ = 1-0) and CCS $J_{N}$ = $8_{7}-7_{6}$ and $7_{7}-6_{6}$ \label{tab:linePara}}
	\tablehead{
		\colhead{Transition} & \colhead{Frequency} & \colhead{Hyperfine Relative} & \colhead{$Su^{2}$} & \colhead{$g_{u}$} & \colhead{$E_{u}/k$} & \colhead{HPBW} \\
		& \colhead{(MHz)} & \colhead{Intensity} & \colhead{(D$^{2}$)} &  & \colhead{(K)} & \colhead{(\arcsec)}
	}
	\startdata
	N${_2}$H$^{+}$ (1$_{\rm 0}$-0$_{\rm 1}$, group 1) & 93171.880 & 1/9 & 11.56 & 3 & 4.47 & 26.54 \\
	N${_2}$H$^{+}$ (1$_{\rm 2}$-0$_{\rm 1}$, group 2) & 93173.700 & 5/9 & 57.80 & 15 & 4.47 & 26.54 \\
	N${_2}$H$^{+}$ (1$_{\rm 1}$-0$_{\rm 1}$, group 3) & 93176.130 & 1/3 & 34.68 & 5 & 4.47 & 26.54\\
	CCS $J_{N}$ = $8_{7}-7_{6}$ & 93870.107 & 1 & 66.11 & 17 & 19.89 & 26.35 \\
	CCS $J_{N}$ = $7_{7}-6_{6}$ & 90686.381 & 1 & 56.88 & 15 & 26.11 & 27.27 \\
	\enddata
	\tablecomments{Column(1): transition; column(2): rest frequency; column(3): Hyperfine (HF) relative intensity; column(4)-(6): the product of the line strength and the square of the electric dipole moment, the upper state degeneracy as well as the upper-level energy from the Cologne Database for Molecular Spectroscopy\footnote{http://www.astro.uni-koeln.de/cdms/}; column(7): half-power beamwidth.}
\end{deluxetable*}

\section{Results and Analyses}\label{sec:Results}

\subsection{Spectra Fitting Results}\label{sec:Spectra Fitting Results}

\begin{figure*}[htbp]
	\centering
	{\includegraphics[width=18cm]{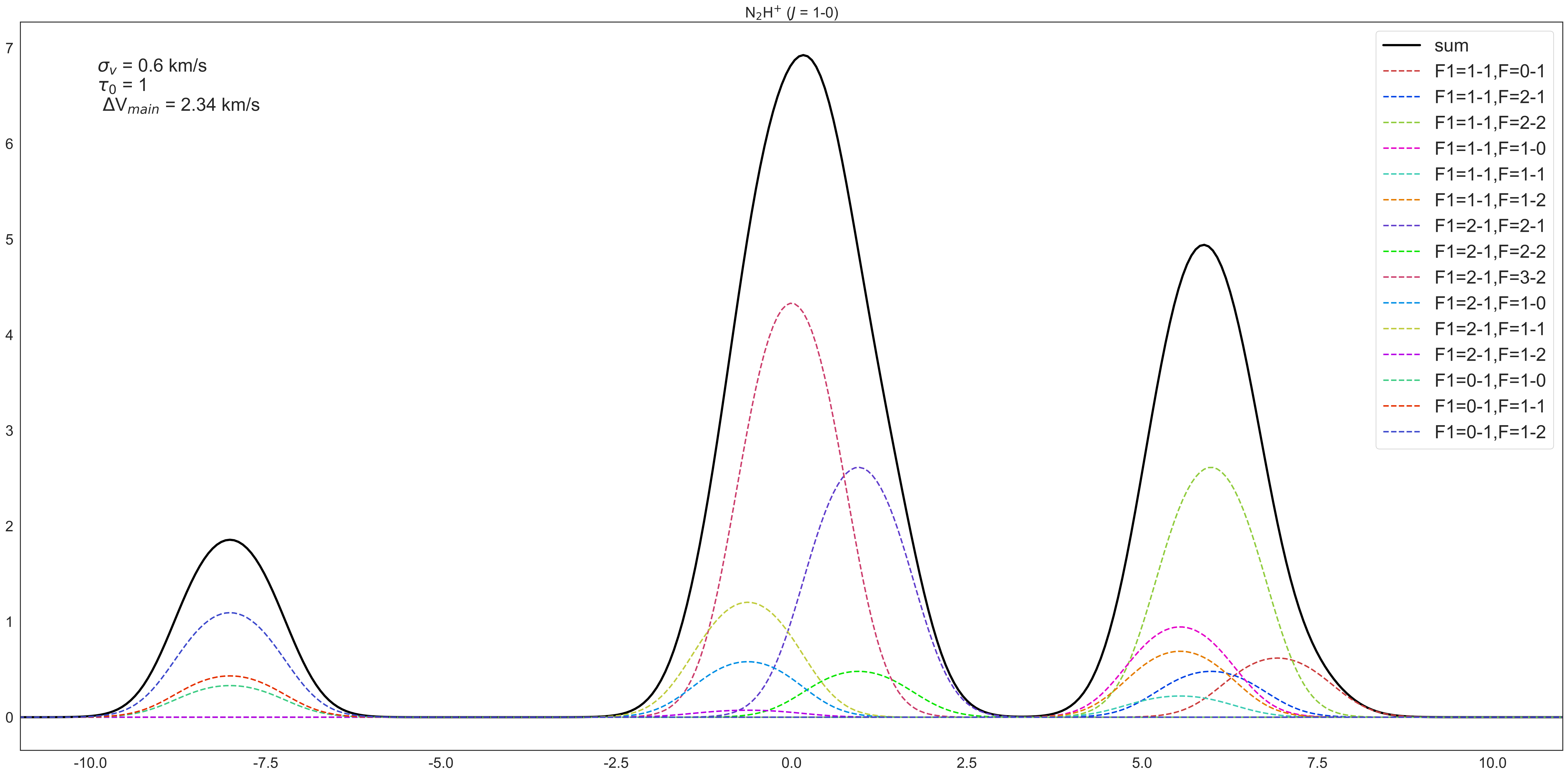}} 
	\caption{Synthetic N$_{2}$H$^{+}$ $J$ = 1$-$0 spectra for an intrinsic velocity dispersion of 0.6 km s$^{-1}$ and a Gaussian line shape.}
	\label{fig:Synthetic_spectra}
\end{figure*}

The Continuum and Line Analysis Single-dish Software (CLASS), which is part of the Grenoble Image and Line Data Analysis Software \footnote{\url{http://http://www.iram.fr/IRAMFR/GILDAS/}}  \citep[GILDAS, e. g.,][]{2000ASPC..217..299G}, is used for data reduction. A first-order polynomial baseline was fitted and subtracted from the averaged spectra for each source, with a velocity resolution of  
$\sim$0.6 $\rm km \, \rm s^{-1}$. 
Among 88 targets, 87 and 33 sources were detected in the N$_{2}$H$^{+}$ $J$ = 1$-$0 and CCS $J_{N}$ = $8_{7}-7_{6}$ lines, respectively. All sources with detection of CCS $J_{N}$ = $8_{7}-7_{6}$ have detection of N$_{2}$H$^{+}$ $J$ = 1$-$0. For the CCS $7_{7}-6_{6}$ transition, we detected emission in 10 out of 82 sources, all of which also exhibited emission in the CCS $J_{N}$ = $8_{7}-7_{6}$ line. The N$_{2}$H$^{+}$ $J$ = 1$-$0 transition line theoretically consists of fifteen hyperfine (HF) components. Our detected N$_{2}$H$^{+}$ lines with relatively broad line width (\textgreater 2 $\rm km \, \rm s^{-1}$) lead to the blending of these fifteen components into three distinct groups (Figure \ref{fig:Synthetic_spectra}), each exhibiting a roughly Gaussian shape \cite[e.g.,][]{2009MNRAS.394..323P,2019A&A...622A..32L}. Thus we tried to fit the N$_{2}$H$^{+}$ ($J$ = 1-0) spectra using three Gaussian profiles. 
For seven sources (G001.00, G028.39, G030.70, G030.81, G032.79, G34.3+0.2, and G097.53) with blending velocity components, we used the "Print area" method \footnote{The "Print area" command in CLASS is a computational tool designed to calculate the integrated intensity of spectral line emission within a user-defined velocity range.} in CLASS to determine the total integrated intensity. The spectra and line parameters from Gaussian fits are showed in Figure \ref{fig:spectrum} and Table \ref{tab:obserUCHII}, respectively. 

\startlongtable
\renewcommand\tabcolsep{7pt} 
\begin{deluxetable*}{llcccccl}
			\tablecaption{Our UC H{\footnotesize II} observational parameters of N$_{2}$H$^{+}$ ($J$ = 1-0) and CCS obtained from Gaussian fits  \label{tab:obserUCHII}}
			\tablewidth{0pt}
			\tablehead{
				\colhead{Object}&
				\colhead{\textbf{Molecule}}&
				\colhead{Total Time}&
				\colhead{Rms}&
				\colhead{$ \int{T_{\rm mb}{\rm{d}} v}$}&
				\colhead{$V_{\rm LSR}$ }&
				\colhead{$\Delta$ V}&
				\colhead{$T_{\rm mb}$}\\
				\colhead{}&
				\colhead{Transition}&
				\colhead{(min)}&
				\colhead{(K)}&
				\colhead{(${\rm K} \, {\rm km} \, {\rm s}^{-1}$)}&
				\colhead{($\rm km \, \rm s^{-1}$)}&
				\colhead{($\rm km \, \rm s^{-1}$)}&
				\colhead{(K)}
			}
			\decimalcolnumbers
			\startdata
			G133.94+01.06 & N$_{2}$H$^{+}$ (1$_{0}$-0$_{1}$, group 1) & 18.11  & 0.021 & 3.14 $\pm$ 0.04  & -54.55 $\pm$ 0.01 & 2.86 $\pm$ 0.04  & 1.03  \\
			& N$_{2}$H$^{+}$ (1$_{2}$-0$_{1}$, group 2) & 18.11  & 0.021 & 13.8 $\pm$ 0.05  & -46.44 $\pm$ 0.01 & 3.51 $\pm$ 0.01  & 3.69  \\
			& N$_{2}$H$^{+}$ (1$_{1}$-0$_{1}$, group 3) & 18.11  & 0.021 & 7.25 $\pm$ 0.05  & -40.7 $\pm$ 0.01  & 3.11 $\pm$ 0.02  & 2.19  \\
			& CCS ($J_{N}$ = $8_{7}-7_{6}$)             & 18.11  & 0.021 & 0.38 $\pm$ 0.05  & -47.02 $\pm$ 0.23 & 3.63 $\pm$ 0.60  & 0.10  \\
			& CCS ($J_{N}$ = $7_{7}-6_{6}$)             &        &       &                  &                   &                  &       \\
			G359.13       & N$_{2}$H$^{+}$ (1$_{0}$-0$_{1}$, group 1) & 3.86   & 0.057 & 0.82 $\pm$ 0.13  & -10.24 $\pm$ 0.25 & 3.18 $\pm$ 0.61  & 0.24  \\
			& N$_{2}$H$^{+}$ (1$_{2}$-0$_{1}$, group 2) & 3.86   & 0.057 & 3.56 $\pm$ 0.15  & -1.93 $\pm$ 0.07  & 3.57 $\pm$ 0.17  & 0.94  \\
			& N$_{2}$H$^{+}$ (1$_{1}$-0$_{1}$, group 3) & 3.86   & 0.057 & 2.29 $\pm$ 0.15  & 3.86 $\pm$ 0.10   & 3.46 $\pm$ 0.28  & 0.62  \\
\enddata
			\tablecomments{
				Column(1): source name; column(2): molecule and transition lines; column(3): total integration time; column(4): the rms noise value; column(5): the integrated line intensities of N$_{2}$H$^{+}$ and CCS, covering all groups of hyperfine (HF) components from Gaussian fitting; 
				column(6): LSR velocity; column(7): line width (FWHM); column(8): main beam brightness temperature for N$_{2}$H$^{+}$ and CCS. For six sources without observed in CCS ($J_{N}$ = $7_{7}-6_{6}$) line  (G133.94+01.06, G009.62+00.19, G012.81-00.19, G34.3+0.2, G033, and G035.14), the line information of them is blank. 
			}
		\end{deluxetable*}


\begin{figure*}[htbp]
	{\includegraphics[width=18cm]{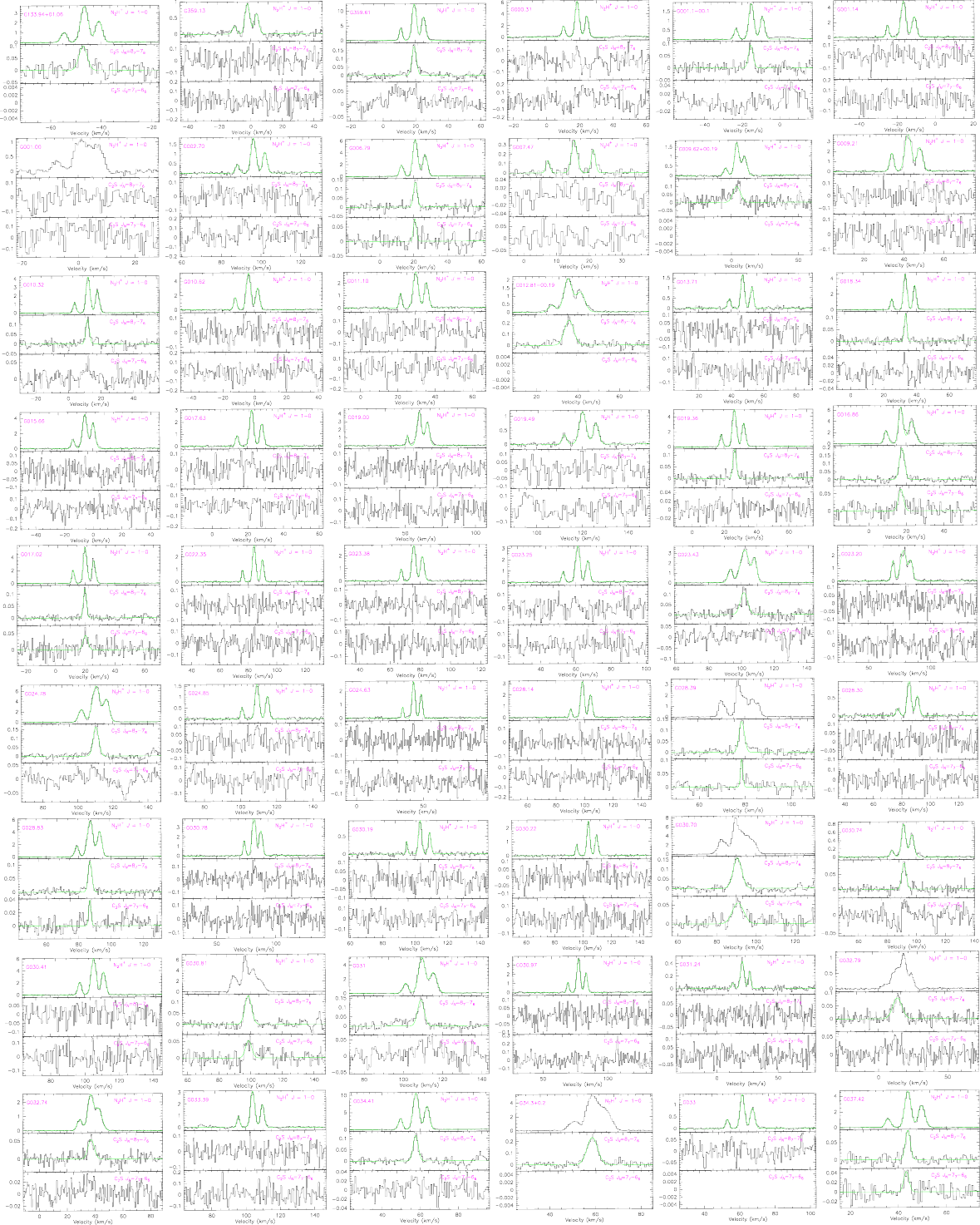}}
	\caption{
	The spectra of N${_2}$H$^{+}$ $J$ = 1$-$0 (top panels), CCS $J_{N}$ = $8_{7}-7_{6}$ (middle panels) and CCS $J_{N}$ = $7_{7}-6_{6}$ (bottom panels) of our UC H{\footnotesize II} sample observed by the IRAM 30 m telescope. 
	For those detections, Green gaussian fit lines were presented. For those six sources that were not observed in CCS $J_{N}$ = $7_{7}-6_{6}$ line, the corresponding panels are blank.
	}
	\label{fig:spectrum}
\end{figure*}

\begin{figure*}[htbp]
	\addtocounter{figure}{-1} 
	{\includegraphics[width=18cm]{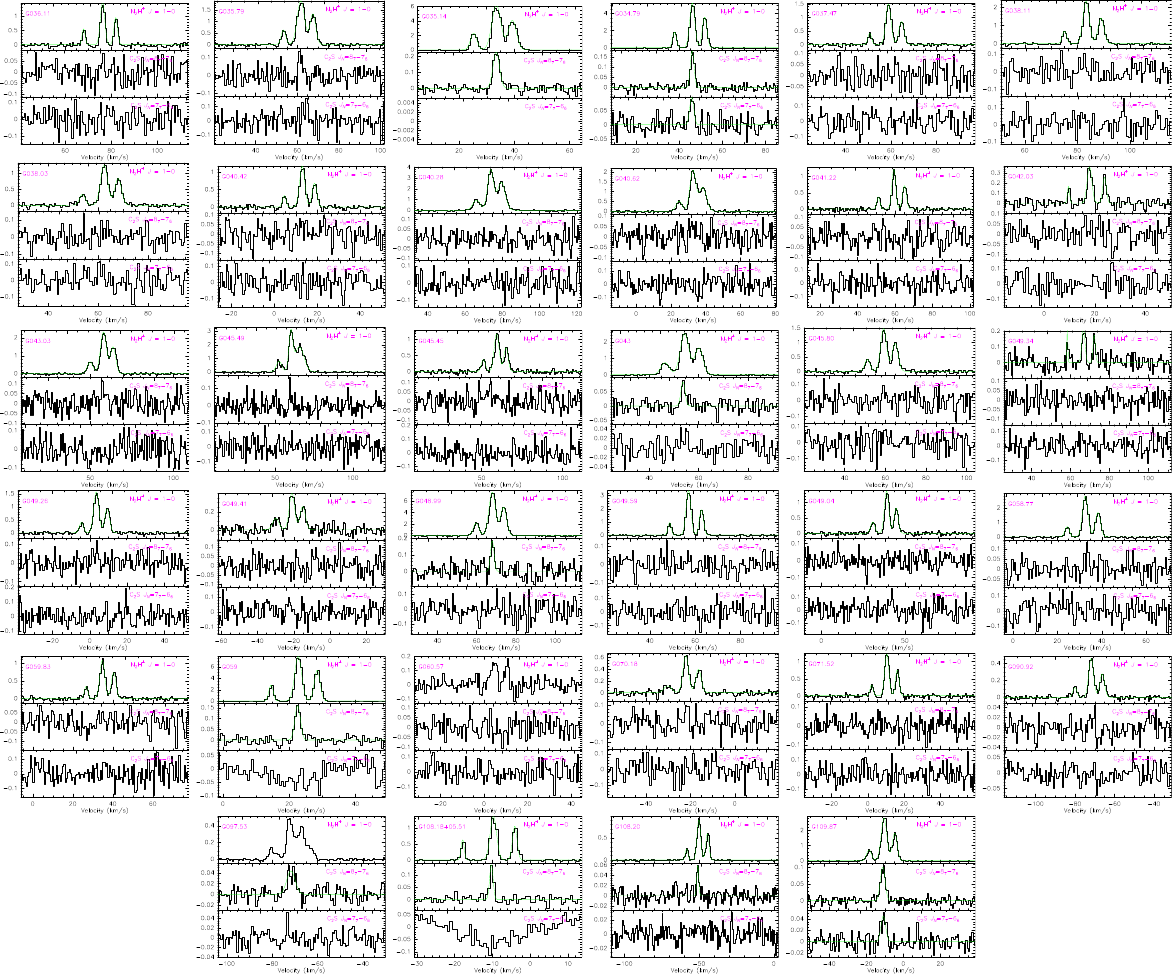}}
	\caption{(Continued.)}
\end{figure*}


\subsection{Line width}\label{sec:Line width}

The linewidth obtained from the Gaussian fit of detected lines provides insights into the state of turbulence and the predominant emission region of the gas \citep[e.g.,][]{2002ApJ...566..945B}. It encompasses both thermal and nonthermal components. The FWHM line width  due to thermal motion ($\Delta V_{\rm T}$) can be estimated using the following formula \cite[e.g.,][]{2003ApJ...586..286L,2014A&A...567A..78L,2018A&A...611A...6T,2023ApJS..266...29Z}: 
\begin{equation}   \label{equ:thermal motion}                                    
	\Delta V_{\rm T} = \sqrt{\frac{8ln2kT_{\rm kin}}{m}},
\end{equation} 
where the molecular mass (m) of the gas is 29 amu and 56 amu for N$_{2}$H$^{+}$ and for CCS, $k$ is the Boltzmann constant, and $T_{\rm kin}$ is the kinetic temperature of the gas. Subsequently, the non-thermal FWHM line width ($\Delta V_{\rm NT}$) can be determined by
\begin{equation}   \label{equ:nonthermal}                                    
	\Delta V_{\rm NT} = (\Delta V_{\rm obs}^{2} - \Delta V_{\rm T}^{2})^{1/2},
\end{equation} 
where $\Delta V_{obs}$ is the observed FWHM line width. 

\begin{figure*}[htbp]
	\centering  
	\subfigure[]{\includegraphics[width=9.3cm]{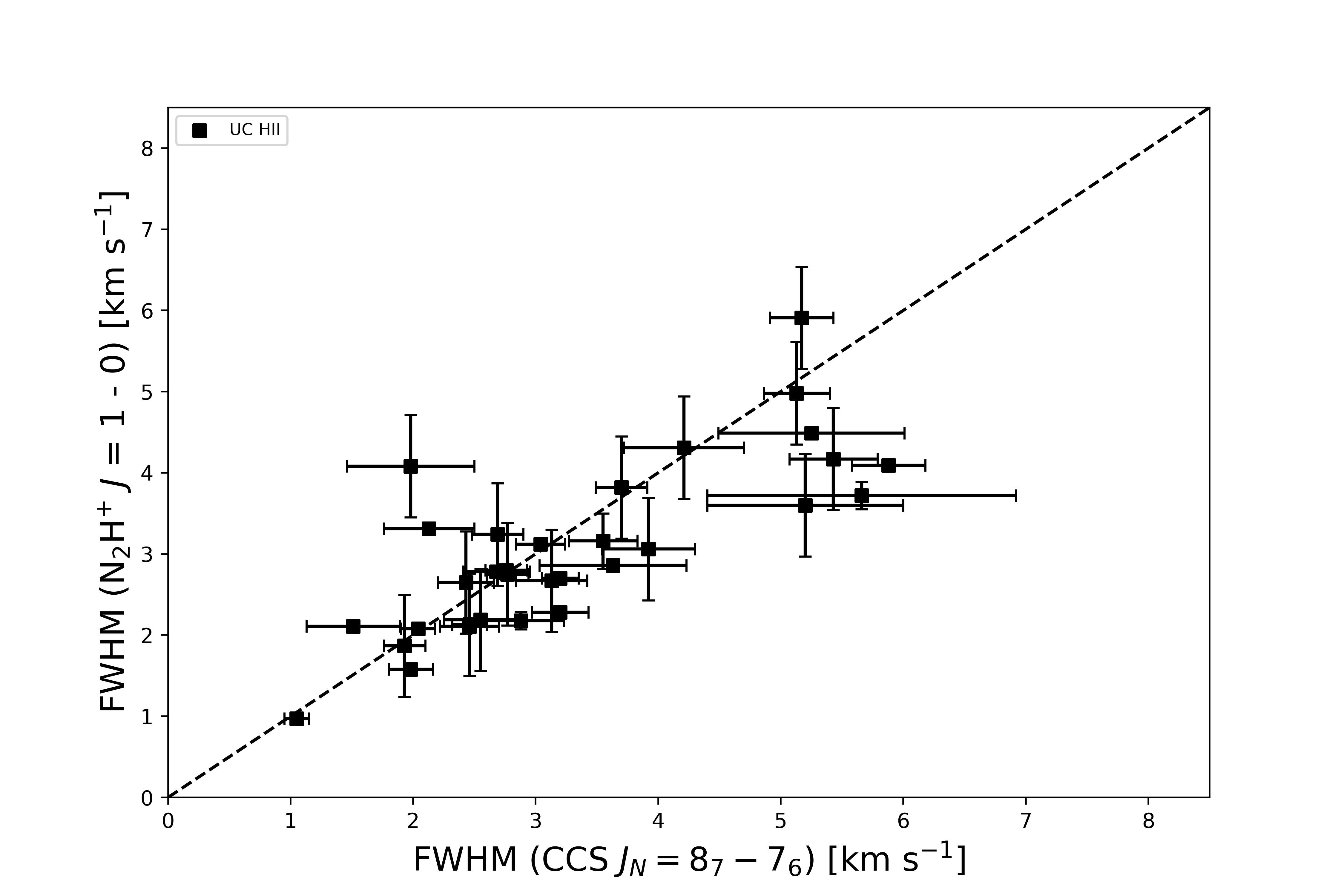}}
	\subfigure[]{\includegraphics[width=9.3cm]{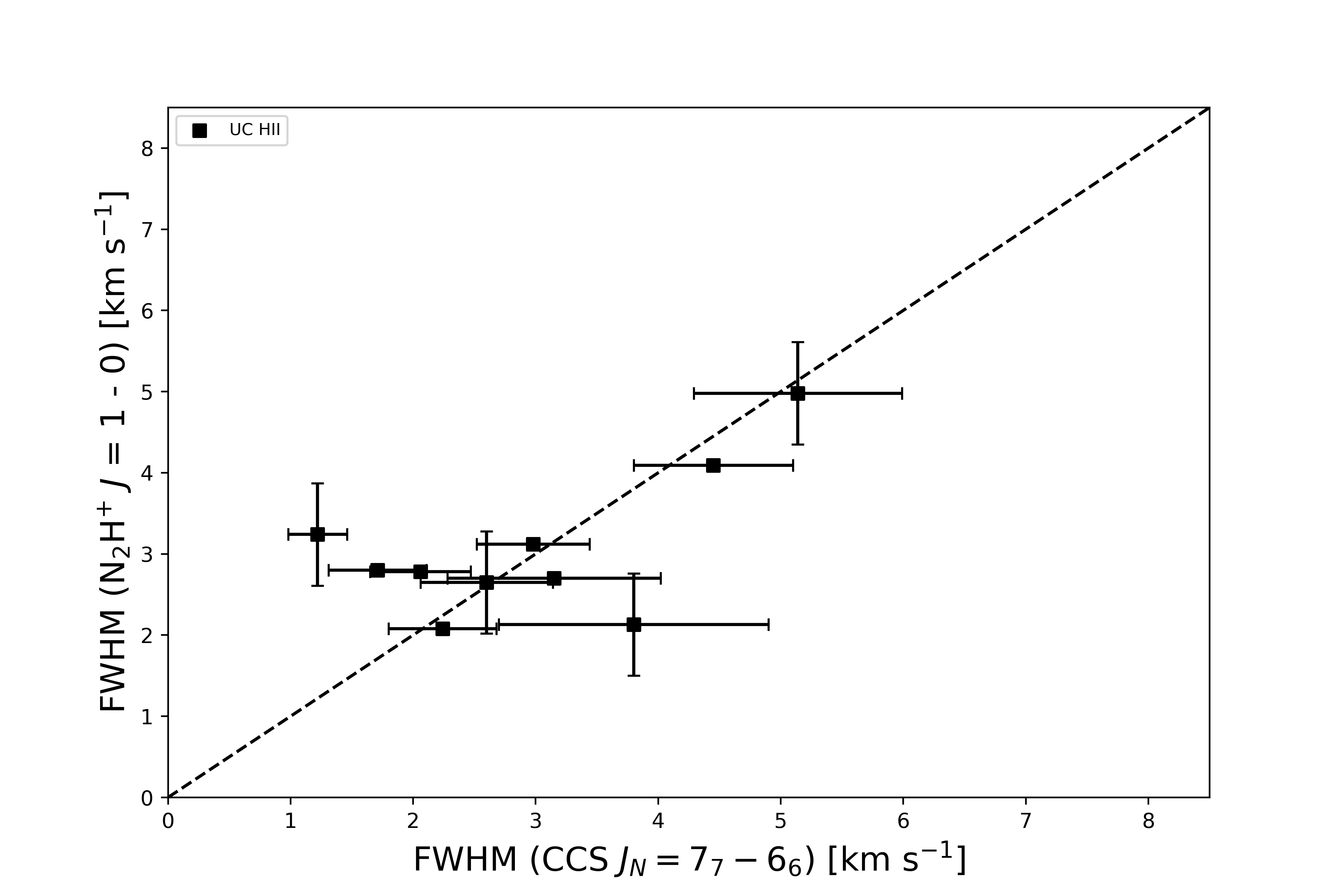}}
	\caption{Two panels show the comparison of the line width between CCS $J_{N}$ = $8_{7}-7_{6}$ and $7_{7}-6_{6}$ and N$_{2}$H$^{+}$ $J$ = 1$-$0. The black dashed line means that both lines have the same line width.
	}
	\label{fig:comparison of line width}
\end{figure*}

By cross-matching our 88 sources with those reported in Reference \citep{2010MNRAS.402.2682H,2011ApJ...741..110D,2011MNRAS.418.1689U,2016ApJ...822...59S,2021ApJS..257...39C}, we obtained the kinetic temperature for each source. These temperatures were estimated from the para-NH$_{3}$ (1, 1) and (2, 2) transitions, with a maximum value of $\sim$40 K. Using the derived kinetic temperatures, we then calculated the line widths for our sample. Our computed results reveal that the thermal line width is negligible, reaching a maximum value of 0.24 km s$^{-1}$ and 0.19 km s$^{-1}$  for N$_{2}$H$^{+}$ and for CCS, constituting less than 10\% of the total line width in our sources (mostly \textgreater 2 km s$^{-1}$). We made a comparison analysis on the line width between CCS lines $J_{N}$ = $8_{7}-7_{6}$ and $7_{7}-6_{6}$ and N$_{2}$H$^{+}$ $J$ = 1$-$0. 
For all sources except five sources (G040.28, G040.62, G045.49, G049.41, and G070.18), the FWHM of N$_{2}$H$^{+}$ $J$ = 1$-$0, $F_{1}$ = 0 - 1 group was taken as the FWHM of N$_{2}$H$^{+}$, as this spectral group comprises components with unresolvable frequencies without blending \citep{2019ApJ...872..154T}. For those five sources,  since the linewidth of group 1 is larger than that of group 2, leading an unreliable FWHM of  N$_{2}$H$^{+}$, we used the FWHM of N$_{2}$H$^{+}$ $J$ = 1$-$0, $F_{1}$ = 1 - 1 group as an upper limit value of their N$_{2}$H$^{+}$ linewidth. However, these five sources do not impact the subsequent analysis, as they lack CCS detections and are thus not included in the comparison presented in Figure \ref{fig:comparison of line width}.
Figure \ref{fig:comparison of line width} shows that the line width of both CCS lines tends to be larger than that of N$_{2}$H$^{+}$ in our UC H{\footnotesize II} sample, i.e., our sources are mostly below the equal-linewidth-line. The mean line width is 2.78, 3.44 and 2.94 km s$^{-1}$ for N$_{2}$H$^{+}$ $J$ = 1$-$0, CCS $J_{N}$ = $8_{7}-7_{6}$ and CCS $J_{N}$ = $7_{7}-6_{6}$, respectively. Analyzing the line width of various HC$_{3}$N transition lines, \cite{2021PASJ...73..467F} contended that inner dense warm regions exhibit more turbulence than outer regions. In this case, our results suggest that CCS is more likely to be present in inner and more dynamically active star-forming regions compared to N$_{2}$H$^{+}$, given that the line width is primarily caused from non-thermal motion, e.g., turbulence. 

\subsection{Physical Parameters of N$_{2}$H$^{+}$ and CCS}\label{sec:Physical parameters}

\subsubsection{Excitation Temperature and Column Density of N$_{2}$H$^{+}$}

To determine precisely the column density of N$_{2}$H$^{+}$, we need to obtain the optical depth of N$_{2}$H$^{+}$. Following the procedure described by \cite{2009MNRAS.394..323P}, the optical depth of N$_{2}$H$^{+}$ can be estimated through the line intensity ratio method. Assuming equal line width for all individual HF components and  under an optically thin condition, the theoretical line intensity ratio of group 1/group 2 (see details in Table \ref{tab:linePara}) should be 0.2 \citep{2015PASP..127..266M}. The optical depth of N$_{2}$H$^{+}$ can be determined with the following formula \cite[e.g.,][]{2021ApJS..257...39C,2024ApJ...971..164C}:
\begin{equation}   \label{equ:optical depth}                                    
	\frac{T_{\rm mb,\;  group \; 1}}{T_{\rm mb, \; group \; 2}} = \frac{1 - e^{-0.2\tau_{2}}}{1 - e^{-\tau_{2}}},
\end{equation} 
where $T_{\rm mb}$ and $\tau_{2}$ are the peak value of the main beam brightness temperature and the optical depth of N$_{2}$H$^{+}$ group 2. We thus used this intensity ratio method to estimate the optical depth of N$_{2}$H$^{+}$ of our sample, except for those 7 sources with blending velocity components in N$_{2}$H$^{+}$. For them, the optical depth of N$_{2}$H$^{+}$ was determined by using HF fitting \cite[the "method" command in CLASS, e.g. ,][]{2021ApJS..257...39C,2024ApJ...971..164C}. These resulting optical depths were then used for our later analyses. To assess the consistency between the two estimation methods, we also estimated the optical depth of remaining 81 sources by using HF fitting method and compared them with those obtained from the intensity ratio (see detail in Appendix \ref{sec: AppendixA}).

And then the excitation temperature of N$_{2}$H$^{+}$ can be calculated with the following equation \cite[e.g.,][]{2023ApJS..264...48W}:
\begin{equation}  \label{equ:excitation}      
	T_{ex} = 4.47/ln\bigg(1+\bigg[\frac{T_{\rm mb, \; group \; 2}}{4.47(1-e^{-\tau_{2}})}+0.236\bigg] \bigg).
\end{equation} 
Finally, we derived the column density of N$_{2}$H$^{+}$ using the following formula \citep{2015PASP..127..266M,2023arXiv231212261Y}:
\begin{equation}\label{equ:total column density}               
	\begin{aligned}
		N &= \bigg(\frac{3h}{8\pi^{3}S\mu^{2}R_{i}}\bigg)\bigg(\frac{Q(T_{ex})}{g_{u}}\bigg)\frac{exp\big(\frac{E_{u}}{kT_{ex}}\big)}{exp\big(\frac{h\nu}{kT_{ex}}\big)-1}\\&\times \frac{\int T_{\rm mb} d v }{J_{\nu}(T_{ex})-J_{\nu}(T_{bg})}\frac{\tau_{2}}{1-exp(-\tau_{2})},
	\end{aligned}
\end{equation}
where $h$ and $\int T_{\rm mb} d v$ are the Planck constant and the integrated line intensity of N$_{2}$H$^{+}$ group 2, respectively, $S$ is the line strength, $\mu$ is the permanent electric dipole moment, $R_{i}$ (= 5/9) is the corresponding theoretical HF relative intensity of N$_{2}$H$^{+}$ group 2, $Q$ is the partition function, $g_{u}$ is the upper state degeneracy, $E_{u}$ is the upper-level energy, $\nu$ is rest frequency. 
$J_{v}(T)$ is  the equivalent temperature of a black body at temperature T \citep{2015PASP..127..266M}:
\begin{equation} \label{Temp:Jv}                                           
	J_{\nu}(T)=\frac{\frac{hv}{k}}{exp({\frac{hv}{kT}})-1}.
\end{equation} 
The derived parameters of N$_{2}$H$^{+}$ of our sample, including the optical depth, the excitation temperature, and the column density, are summarized in Table \ref{tab:resultUCHII}. These resulting excitation temperature and column density may be lower limits, since the beam-filling factor was not considered here.

\startlongtable
\renewcommand\tabcolsep{6.5pt} 
		\begin{deluxetable*}{lcccccc}
			\tablecaption{Derived parameters of CCS and N$_{2}$H$^{+}$ in our UC H{\footnotesize II} sample \label{tab:resultUCHII}}
			\tablehead{
				\multirow{2}{*}{source} & 
				\multicolumn{3}{c}{N$_{2}$H$^{+}$}                     & 
				\multicolumn{2}{c}{CCS}                & 
				\multirow{2}{*}{$\frac{N (\rm N_{2}H^{+})}{N (\rm CCS)}$} 
				\\ \cmidrule(lr){2-4} \cmidrule(lr){5-6}
				\colhead{}&
				\colhead{$\tau$}           & 
				\colhead{$T_{ex}$}          & 
				\colhead{$N$}             &  
				\colhead{T$_{rot}$}             & 
				\colhead{$N$}             &                                    
				\colhead{}                         
				\\        
				\colhead{}&                 
				\colhead{}& 
				\colhead{(K)}               & 
				\colhead{($\times$ 10$^{13}$ cm$^{-2}$)}           &     
				\colhead{(K)}            &  
				\colhead{($\times$ 10$^{11}$ cm$^{-2}$)}            &                                  
				\colhead{}
				}
			\decimalcolnumbers
			\startdata
			G133.94+01.06 & 0.92 $\pm$ 0.04 & 9.28 $\pm$ 0.69  & 2.70 $\pm$ 0.31  & 11.93 $\pm$ 5.72$^{*}$ & 3.23  $\pm$ 0.75$^{*}$  & 83.83  $\pm$ 21.62  \\
			G359.13       & 0.65 $\pm$ 0.10 & 4.94 $\pm$ 0.33  & 0.76 $\pm$ 0.32  & ...                    & ...                     & ...                 \\
			G359.61       & 1.55 $\pm$ 0.05 & 18.90 $\pm$ 1.66 & 15.94 $\pm$ 2.26 & 11.93 $\pm$ 5.72$^{*}$ & 5.01  $\pm$ 0.76$^{*}$  & 318.25  $\pm$ 65.90 \\
			G000.31       & 1.12 $\pm$ 0.06 & 12.36 $\pm$ 1.14 & 4.51 $\pm$ 0.68  & ...                    & ...                     & ...                 \\
			G001.1-00.1   & 1.02 $\pm$ 0.06 & 5.96 $\pm$ 0.36  & 1.25 $\pm$ 0.16  & 11.93 $\pm$ 5.72$^{*}$ & 2.38  $\pm$ 0.49$^{*}$  & 52.75  $\pm$ 12.90  \\
			G001.14       & 1.50 $\pm$ 0.06 & 9.37 $\pm$ 0.71  & 3.84 $\pm$ 0.42  & ...                    & ...                     & ...                 \\
			G001.00       & 1.85 $\pm$ 0.11 & 4.22 $\pm$ 0.23  & 2.90 $\pm$ 0.51  & ...                    & ...                     & ...                 \\
			G002.70       & 0.51 $\pm$ 0.06 & 7.60 $\pm$ 0.61  & 0.90 $\pm$ 0.10  & ...                    & ...                     & ...                 \\
			G006.79       & 1.17 $\pm$ 0.04 & 12.17 $\pm$ 0.95 & 5.53 $\pm$ 0.69  & 11.22  $\pm$ 3.89      & 3.42  $\pm$ 0.61        & 161.74  $\pm$ 35.01 \\
			G007.47       & 2.01 $\pm$ 0.11 & 3.20 $\pm$ 0.19  & 0.73 $\pm$ 0.17  & ...                    & ...                     & ...                 \\
			G009.62+00.19 & 0.44 $\pm$ 0.05 & 7.93 $\pm$ 0.79  & 1.22 $\pm$ 0.14  & 11.93 $\pm$ 5.72$^{*}$ & 5.18  $\pm$ 1.37$^{*}$  & 23.61  $\pm$ 6.79   \\
			G009.21       & 3.00 $\pm$ 0.08 & 7.09 $\pm$ 0.47  & 6.56 $\pm$ 0.91  & ...                    & ...                     & ...                 \\
			G010.32       & 1.03 $\pm$ 0.04 & 9.72 $\pm$ 0.71  & 2.77 $\pm$ 0.41  & 11.93 $\pm$ 5.72$^{*}$ & 2.89  $\pm$ 0.54$^{*}$  & 96.07  $\pm$ 22.97  \\
			G010.62       & 1.52 $\pm$ 0.06 & 9.73 $\pm$ 0.74  & 4.52 $\pm$ 0.50  & ...                    & ...                     & ...                 \\
			G011.10       & 2.29 $\pm$ 0.08 & 6.23 $\pm$ 0.41  & 3.29 $\pm$ 0.37  & ...                    & ...                     & ...                 \\
			G012.81-00.19 & 0.10 $\pm$ 0.04 & 5.04 $\pm$ 7.53  & 1.49 $\pm$ 0.21  & 11.93 $\pm$ 5.72$^{*}$ & 11.12  $\pm$ 1.71$^{*}$ & 13.39  $\pm$ 2.78   \\
			G013.71       & 0.93 $\pm$ 0.07 & 5.73 $\pm$ 0.39  & 0.96 $\pm$ 0.16  & ...                    & ...                     & ...                 \\
			G018.34       & 1.24 $\pm$ 0.05 & 9.52 $\pm$ 0.79  & 2.89 $\pm$ 0.44  & 11.93 $\pm$ 5.72$^{*}$ & 2.04  $\pm$ 0.37$^{*}$  & 141.87  $\pm$ 33.70 \\
			G015.66       & 0.73 $\pm$ 0.05 & 11.86 $\pm$ 0.93 & 3.92 $\pm$ 0.76  & ...                    & ...                     & ...                 \\
			G017.63       & 1.10 $\pm$ 0.05 & 7.62 $\pm$ 0.54  & 1.83 $\pm$ 0.20  & ...                    & ...                     & ...                 \\
			G019.00       & 1.03 $\pm$ 0.05 & 9.65 $\pm$ 0.89  & 3.63 $\pm$ 0.43  & ...                    & ...                     & ...                 \\
			G019.49       & 1.07 $\pm$ 0.08 & 4.73 $\pm$ 0.27  & 1.06 $\pm$ 0.23  & ...                    & ...                     & ...                 \\
			G019.36       & 1.42 $\pm$ 0.05 & 9.70 $\pm$ 0.67  & 3.86 $\pm$ 0.41  & 11.93 $\pm$ 5.72$^{*}$ & 2.72  $\pm$ 0.53$^{*}$  & 142.22  $\pm$ 31.43 \\
			G016.86       & 1.70 $\pm$ 0.05 & 11.05 $\pm$ 0.82 & 7.08 $\pm$ 0.77  & 7.16  $\pm$ 1.61       & 9.27  $\pm$ 1.35        & 76.37  $\pm$ 13.86  \\
			G017.02       & 1.88 $\pm$ 0.05 & 9.70 $\pm$ 0.69  & 5.32 $\pm$ 0.74  & 20.76  $\pm$ 15.98     & 2.53  $\pm$ 0.41        & 210.04  $\pm$ 44.84 \\
			G022.35       & 1.38 $\pm$ 0.06 & 6.98 $\pm$ 0.46  & 1.96 $\pm$ 0.21  & ...                    & ...                     & ...                 \\
			G023.38       & 1.51 $\pm$ 0.06 & 6.67 $\pm$ 0.44  & 2.38 $\pm$ 0.27  & ...                    & ...                     & ...                 \\
			G023.25       & 1.09 $\pm$ 0.06 & 7.96 $\pm$ 0.39  & 2.05 $\pm$ 0.22  & ...                    & ...                     & ...                 \\
			G023.43       & 2.19 $\pm$ 0.06 & 6.82 $\pm$ 0.41  & 6.29 $\pm$ 0.84  & 11.93 $\pm$ 5.72$^{*}$ & 5.09  $\pm$ 1.02$^{*}$  & 123.42  $\pm$ 29.67 \\
			G023.20       & 5.24 $\pm$ 0.11 & 5.41 $\pm$ 0.30  & 10.36 $\pm$ 1.20 & ...                    & ...                     & ...                 \\
			G024.78       & 1.62 $\pm$ 0.05 & 11.87 $\pm$ 0.89 & 10.36 $\pm$ 1.47 & 11.93 $\pm$ 5.72$^{*}$ & 5.52  $\pm$ 0.81$^{*}$  & 187.82  $\pm$ 38.23 \\
			G024.85       & 1.45 $\pm$ 0.07 & 5.05 $\pm$ 0.29  & 1.25 $\pm$ 0.19  & ...                    & ...                     & ...                 \\
			G024.63       & 1.13 $\pm$ 0.06 & 7.04 $\pm$ 0.48  & 1.66 $\pm$ 0.18  & ...                    & ...                     & ...                 \\
			G028.14       & 0.45 $\pm$ 0.04 & 11.93 $\pm$ 1.14 & 1.70 $\pm$ 0.48  & ...                    & ...                     & ...                 \\
			G028.39       & 3.48 $\pm$ 0.08 & 6.03 $\pm$ 0.32  & 7.07 $\pm$ 0.98  & 7.34  $\pm$ 1.40       & 5.41  $\pm$ 0.81        & 130.82  $\pm$ 26.69 \\
			G028.30       & 0.08 $\pm$ 0.07 & 3.86 $\pm$ 6.81  & 0.68 $\pm$ 0.10  & ...                    & ...                     & ...                 \\
			G028.83       & 1.21 $\pm$ 0.04 & 11.44 $\pm$ 0.80 & 6.02 $\pm$ 0.71  & 4.88  $\pm$ 0.59       & 10.69  $\pm$ 1.72       & 56.34  $\pm$ 11.22  \\
			G030.78       & 2.13 $\pm$ 0.07 & 7.68 $\pm$ 0.53  & 5.08 $\pm$ 0.54  & ...                    & ...                     & ...                 \\
			G030.19       & 2.02 $\pm$ 0.10 & 3.99 $\pm$ 0.17  & 1.06 $\pm$ 0.18  & ...                    & ...                     & ...                 \\
			G030.22       & 1.33 $\pm$ 0.06 & 6.45 $\pm$ 0.41  & 1.60 $\pm$ 0.18  & ...                    & ...                     & ...                 \\
			G030.70       & 3.73 $\pm$ 0.08 & 9.01 $\pm$ 0.65  & 22.54 $\pm$ 2.84 & 7.67  $\pm$ 1.38       & 10.90  $\pm$ 1.60       & 206.82  $\pm$ 39.98 \\
			G030.74       & 0.03 $\pm$ 0.04 & 3.66 $\pm$ 0.38  & 0.76 $\pm$ 0.10  & 11.93 $\pm$ 5.72$^{*}$ & 2.72  $\pm$ 0.44$^{*}$  & 27.91  $\pm$ 5.90   \\
			G030.41       & 1.69 $\pm$ 0.06 & 10.59 $\pm$ 0.81 & 6.29 $\pm$ 0.69  & ...                    & ...                     & ...                 \\
			G030.81       & 4.82 $\pm$ 0.09 & 7.84 $\pm$ 0.50  & 24.08 $\pm$ 2.49 & 6.08  $\pm$ 1.03       & 16.60  $\pm$ 2.41       & 145.09  $\pm$ 25.89 \\
			G031          & 0.93 $\pm$ 0.04 & 11.51 $\pm$ 0.91 & 4.81 $\pm$ 0.77  & 11.93 $\pm$ 5.72$^{*}$ & 4.92  $\pm$ 0.83$^{*}$  & 97.67  $\pm$ 22.80  \\
			G030.97       & 1.17 $\pm$ 0.05 & 8.44 $\pm$ 0.75  & 2.50 $\pm$ 0.35  & ...                    & ...                     & ...                 \\
			G031.24       & 0.11 $\pm$ 0.09 & 7.04 $\pm$ 0.08  & 0.20 $\pm$ 0.10  & ...                    & ...                     & ...                 \\
			G032.79       & 0.10 $\pm$ 0.18 & 3.49 $\pm$ 0.09  & 1.91 $\pm$ 0.31  & 11.93 $\pm$ 5.72$^{*}$ & 5.77  $\pm$ 1.00$^{*}$  & 33.05  $\pm$ 7.83   \\
			G032.74       & 1.32 $\pm$ 0.05 & 6.37 $\pm$ 0.44  & 3.39 $\pm$ 0.36  & 11.93 $\pm$ 5.72$^{*}$ & 2.55  $\pm$ 0.51$^{*}$  & 133.15  $\pm$ 30.18 \\
			G033.39       & 1.22 $\pm$ 0.05 & 7.47 $\pm$ 0.52  & 2.11 $\pm$ 0.24  & ...                    & ...                     & ...                 \\
			G034.41       & 1.12 $\pm$ 0.04 & 19.02 $\pm$ 1.79 & 12.10 $\pm$ 2.29 & 11.93 $\pm$ 5.72$^{*}$ & 3.23  $\pm$ 0.58$^{*}$  & 375.03  $\pm$ 97.66 \\
			G34.3+0.2     & 0.54 $\pm$ 0.04 & 14.64 $\pm$ 1.23 & 6.22 $\pm$ 1.74  & 11.93 $\pm$ 5.72$^{*}$ & 10.61  $\pm$ 1.57$^{*}$ & 58.61  $\pm$ 18.57  \\
			G033          & 0.79 $\pm$ 0.05 & 5.62 $\pm$ 0.38  & 0.90 $\pm$ 0.15  & ...                    & ...                     & ...                 \\
			G037.42       & 0.52 $\pm$ 0.04 & 14.75 $\pm$ 1.30 & 3.15 $\pm$ 0.90  & 18.83  $\pm$ 10.81     & 1.72  $\pm$ 0.33        & 183.50  $\pm$ 62.97 \\
			G036.11       & 1.93 $\pm$ 0.08 & 4.69 $\pm$ 0.24  & 1.27 $\pm$ 0.18  & ...                    & ...                     & ...                 \\
			G035.79       & 1.52 $\pm$ 0.06 & 5.42 $\pm$ 0.30  & 2.22 $\pm$ 0.28  & ...                    & ...                     & ...                 \\
			G035.14       & 1.88 $\pm$ 0.05 & 10.42 $\pm$ 0.76 & 7.85 $\pm$ 1.14  & 11.93 $\pm$ 5.72$^{*}$ & 5.60  $\pm$ 0.90$^{*}$  & 140.12  $\pm$ 30.34 \\
			G034.79       & 1.67 $\pm$ 0.05 & 9.53 $\pm$ 0.67  & 4.36 $\pm$ 0.46  & 24.65  $\pm$ 17.70     & 3.23  $\pm$ 0.49        & 135.02  $\pm$ 24.84 \\
			G037.47       & 0.87 $\pm$ 0.06 & 5.58 $\pm$ 0.36  & 0.88 $\pm$ 0.15  & ...                    & ...                     & ...                 \\
			G038.11       & 1.10 $\pm$ 0.06 & 6.55 $\pm$ 0.43  & 1.50 $\pm$ 0.17  & ...                    & ...                     & ...                 \\
			G038.03       & 0.86 $\pm$ 0.07 & 5.14 $\pm$ 0.34  & 1.01 $\pm$ 0.21  & ...                    & ...                     & ...                 \\
			G040.42       & 0.95 $\pm$ 0.07 & 4.89 $\pm$ 0.27  & 0.81 $\pm$ 0.18  & ...                    & ...                     & ...                 \\
			G040.28       & 0.55 $\pm$ 0.05 & 11.93 $\pm$ 1.07 & 3.44 $\pm$ 0.78  & ...                    & ...                     & ...                 \\
			G040.62       & 0.58 $\pm$ 0.05 & 7.76 $\pm$ 0.70  & 1.64 $\pm$ 0.19  & ...                    & ...                     & ...                 \\
			G041.22       & 1.23 $\pm$ 0.07 & 4.66 $\pm$ 0.25  & 0.87 $\pm$ 0.16  & ...                    & ...                     & ...                 \\
			G042.03       & 2.97 $\pm$ 0.17 & 3.19 $\pm$ 0.09  & 1.11 $\pm$ 0.27  & ...                    & ...                     & ...                 \\
			G043.03       & 0.94 $\pm$ 0.05 & 6.86 $\pm$ 0.47  & 2.14 $\pm$ 0.24  & ...                    & ...                     & ...                 \\
			G045.49       & 0.92 $\pm$ 0.05 & 7.73 $\pm$ 0.58  & 2.87 $\pm$ 0.32  & ...                    & ...                     & ...                 \\
			G045.45       & 1.32 $\pm$ 0.08 & 4.55 $\pm$ 0.24  & 1.33 $\pm$ 0.25  & ...                    & ...                     & ...                 \\
			G043          & 1.08 $\pm$ 0.05 & 6.82 $\pm$ 0.41  & 2.45 $\pm$ 0.34  & 11.93 $\pm$ 5.72$^{*}$ & 1.53  $\pm$ 0.41$^{*}$  & 160.48  $\pm$ 48.24 \\
			G045.80       & 0.89 $\pm$ 0.06 & 5.52 $\pm$ 0.36  & 1.22 $\pm$ 0.21  & ...                    & ...                     & ...                 \\
			G049.34       & 1.87 $\pm$ 0.41 & 2.99 $\pm$ 0.07  & 6.34 $\pm$ 2.56  & ...                    & ...                     & ...                 \\
			G049.26       & 0.57 $\pm$ 0.06 & 6.66 $\pm$ 0.50  & 0.90 $\pm$ 0.13  & ...                    & ...                     & ...                 \\
			G049.41       & 1.14 $\pm$ 0.14 & 3.43 $\pm$ 0.14  & 0.85 $\pm$ 0.46  & ...                    & ...                     & ...                 \\
			G048.99       & 1.05 $\pm$ 0.05 & 14.75 $\pm$ 1.27 & 8.29 $\pm$ 1.29  & 11.93 $\pm$ 5.72$^{*}$ & 3.57  $\pm$ 0.95$^{*}$  & 232.41  $\pm$ 71.78 \\
			G049.59       & 0.93 $\pm$ 0.05 & 8.65 $\pm$ 0.64  & 1.68 $\pm$ 0.19  & ...                    & ...                     & ...                 \\
			G049.04       & 0.76 $\pm$ 0.07 & 5.58 $\pm$ 0.38  & 0.90 $\pm$ 0.18  & ...                    & ...                     & ...                 \\
			G058.77       & 0.59 $\pm$ 0.05 & 9.40 $\pm$ 0.83  & 1.50 $\pm$ 0.22  & ...                    & ...                     & ...                 \\
			G059.83       & 1.30 $\pm$ 0.07 & 4.43 $\pm$ 0.22  & 1.05 $\pm$ 0.20  & ...                    & ...                     & ...                 \\
			G059          & 1.73 $\pm$ 0.05 & 12.67 $\pm$ 1.01 & 6.36 $\pm$ 0.72  & 11.93 $\pm$ 5.72$^{*}$ & 2.80  $\pm$ 0.53$^{*}$  & 226.99  $\pm$ 50.42 \\
			G060.57       & ...             & ...              & ...              & ...                    & ...                     & ...                 \\
			G070.18       & 0.10 $\pm$ 0.08 & 3.48 $\pm$ 4.94  & 0.60 $\pm$ 0.10  & ...                    & ...                     & ...                 \\
			G071.52       & 0.73 $\pm$ 0.07 & 5.36 $\pm$ 0.35  & 0.69 $\pm$ 0.16  & ...                    & ...                     & ...                 \\
			G090.92       & 0.47 $\pm$ 0.06 & 4.15 $\pm$ 0.19  & 0.34 $\pm$ 0.21  & ...                    & ...                     & ...                 \\
			G097.53       & 0.66 $\pm$ 0.06 & 3.77 $\pm$ 0.09  & 0.72 $\pm$ 0.30  & 11.93 $\pm$ 5.72$^{*}$ & 1.61  $\pm$ 0.33$^{*}$  & 44.36  $\pm$ 20.95  \\
			G108.18+05.51 & 2.91 $\pm$ 0.08 & 4.37 $\pm$ 0.17  & 1.53 $\pm$ 0.19  & 11.93 $\pm$ 5.72$^{*}$ & 1.27  $\pm$ 0.21$^{*}$  & 120.51  $\pm$ 24.87 \\
			G108.20       & 0.94 $\pm$ 0.05 & 4.41 $\pm$ 0.19  & 0.73 $\pm$ 0.14  & 11.93 $\pm$ 5.72$^{*}$ & 0.85  $\pm$ 0.25$^{*}$  & 85.53  $\pm$ 30.71  \\
			G109.87       & 0.64 $\pm$ 0.04 & 9.01 $\pm$ 0.88  & 1.88 $\pm$ 0.22  & 10.70  $\pm$ 2.85      & 2.98  $\pm$ 0.48        & 63.03  $\pm$ 12.55 \\
			\enddata
			\tablecomments{
				Column(1): source name; column(2)-(4): the optical depth, the excitation temperature, and the column density of N$_{2}$H$^{+}$; column(5)-(6): the rotational temperature and the column density of CCS. T$_{rot}$ and $N$ indicated by * represent the average value of T$_{rot}$ and the corresponding column density of CCS; column(7): the column density ratio of N$_{2}$H$^{+}$ and CCS.
			}
		\end{deluxetable*}

\subsubsection{Rotational Temperature and Column Density of CCS}

\begin{figure*}[htbp]
	\graphicspath{{figures//}}
	\centering  
	{\includegraphics[width=4.0cm]{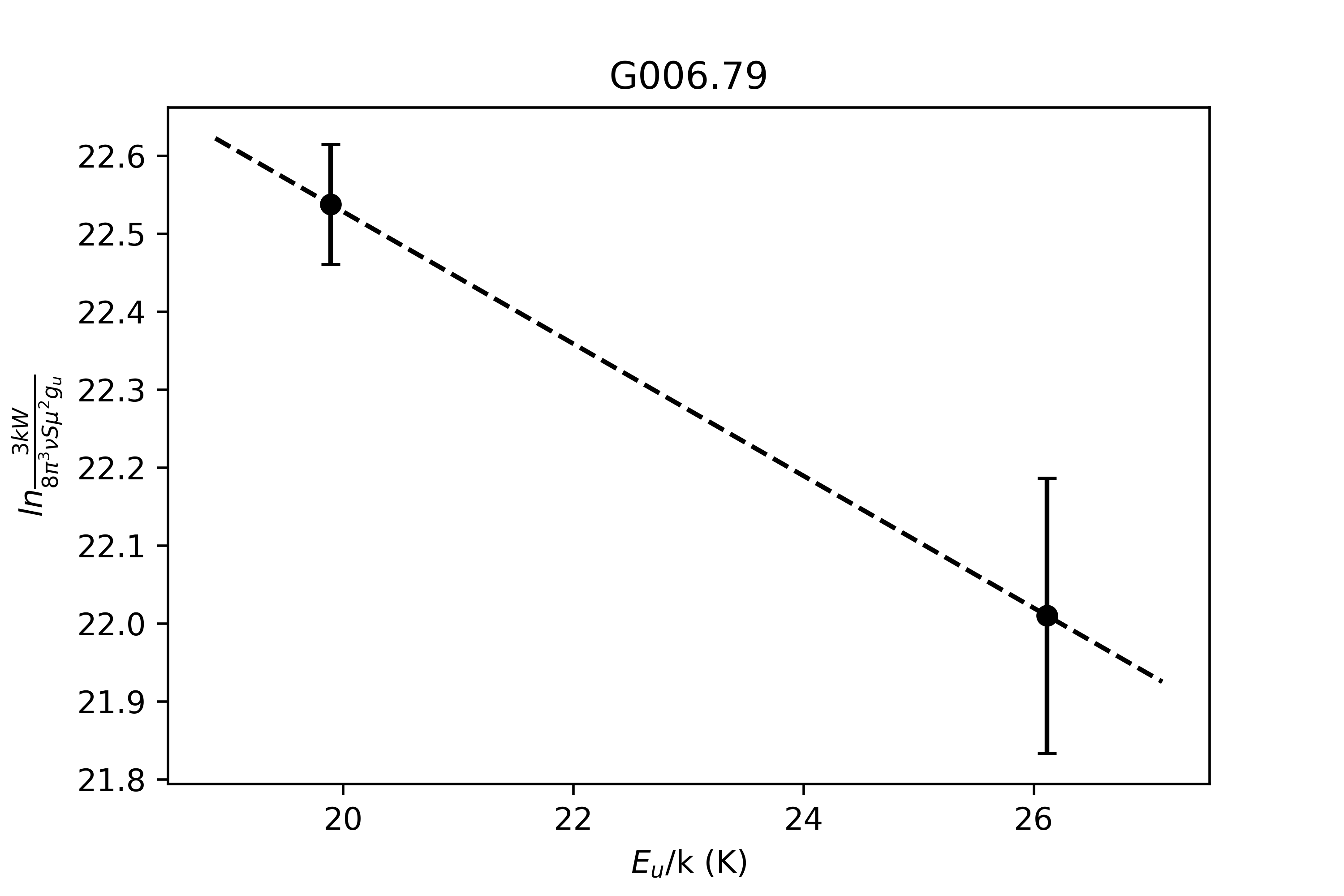}}
	{\includegraphics[width=4.0cm]{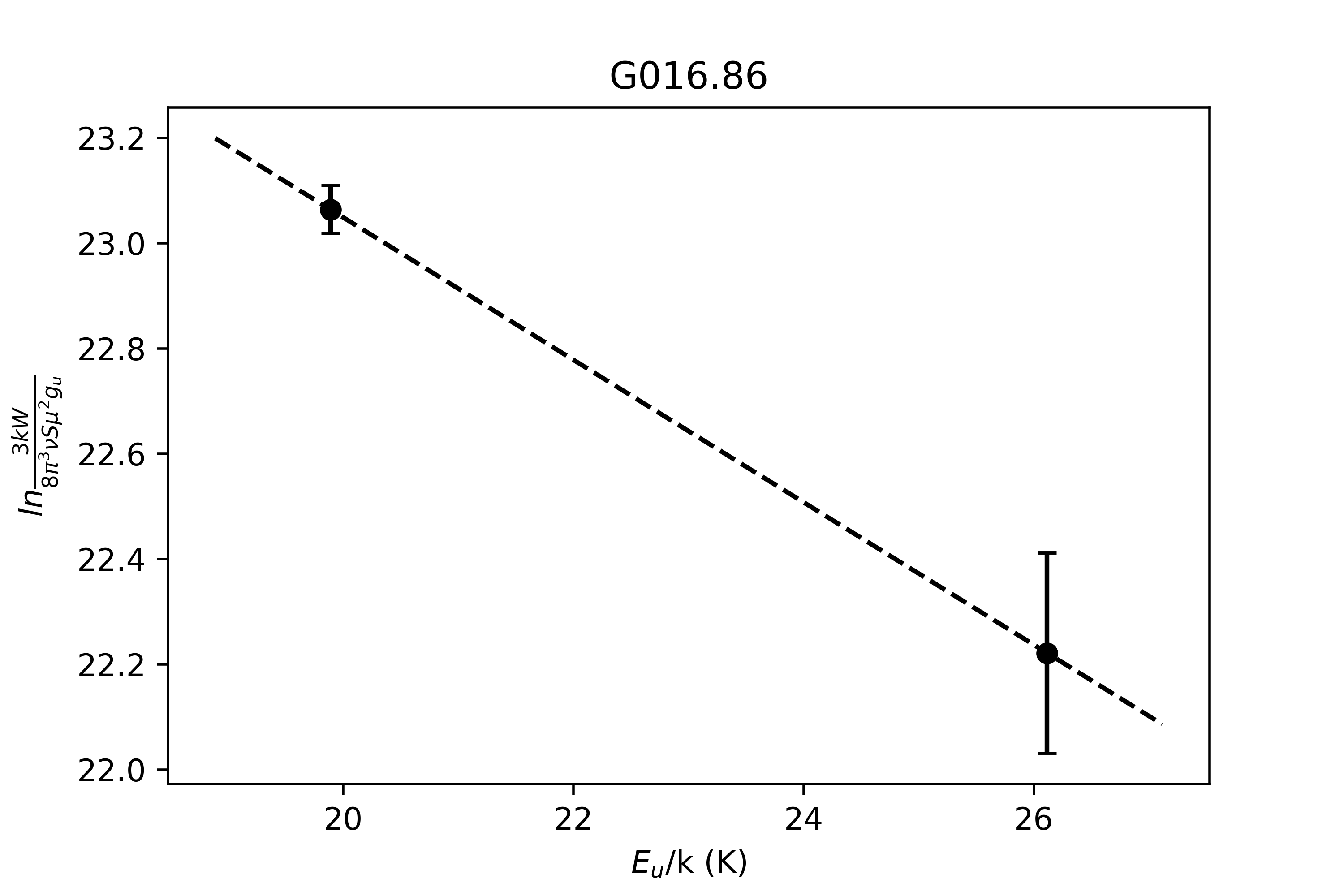}}
	{\includegraphics[width=4.0cm]{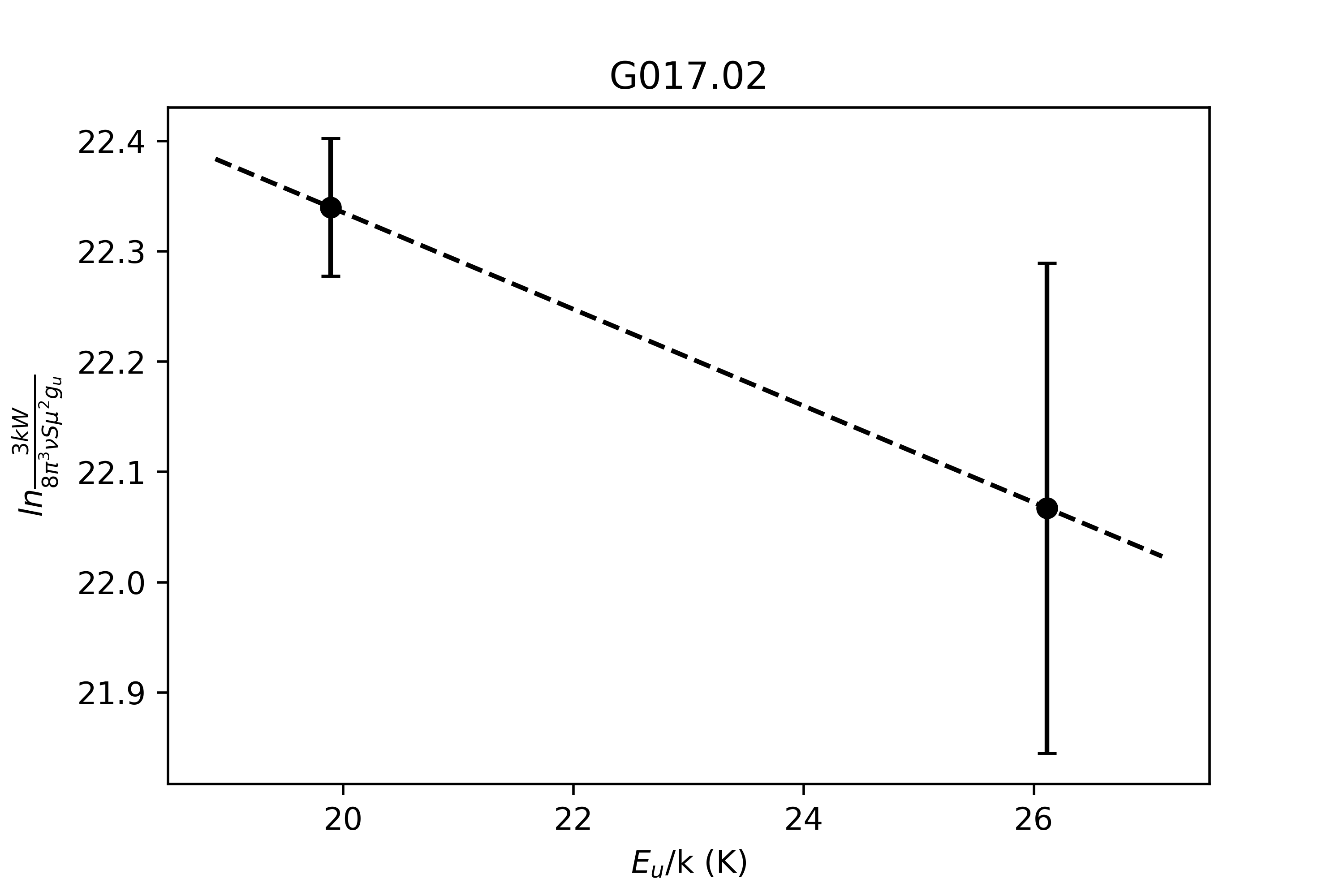}}
	{\includegraphics[width=4.0cm]{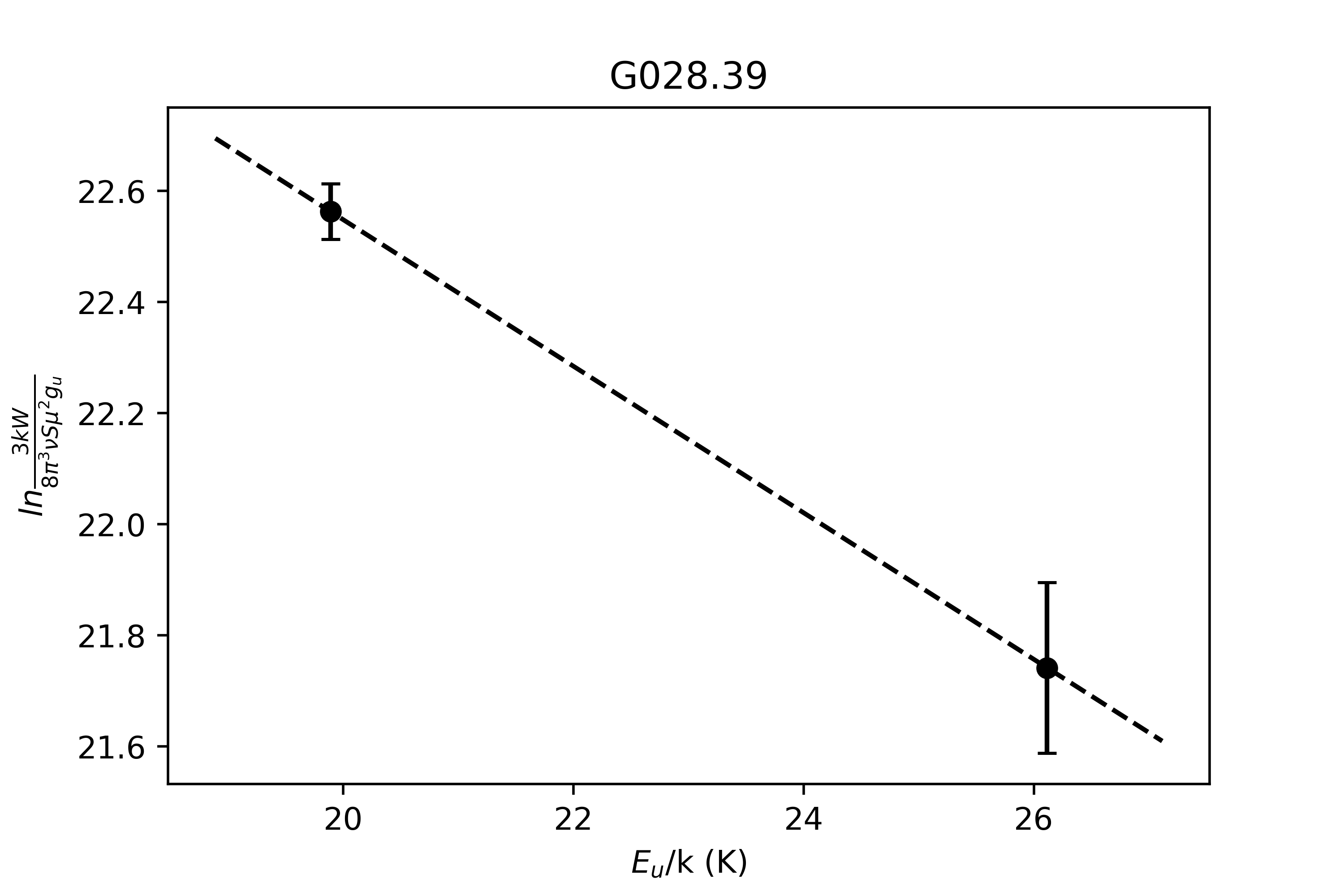}}
	{\includegraphics[width=4.0cm]{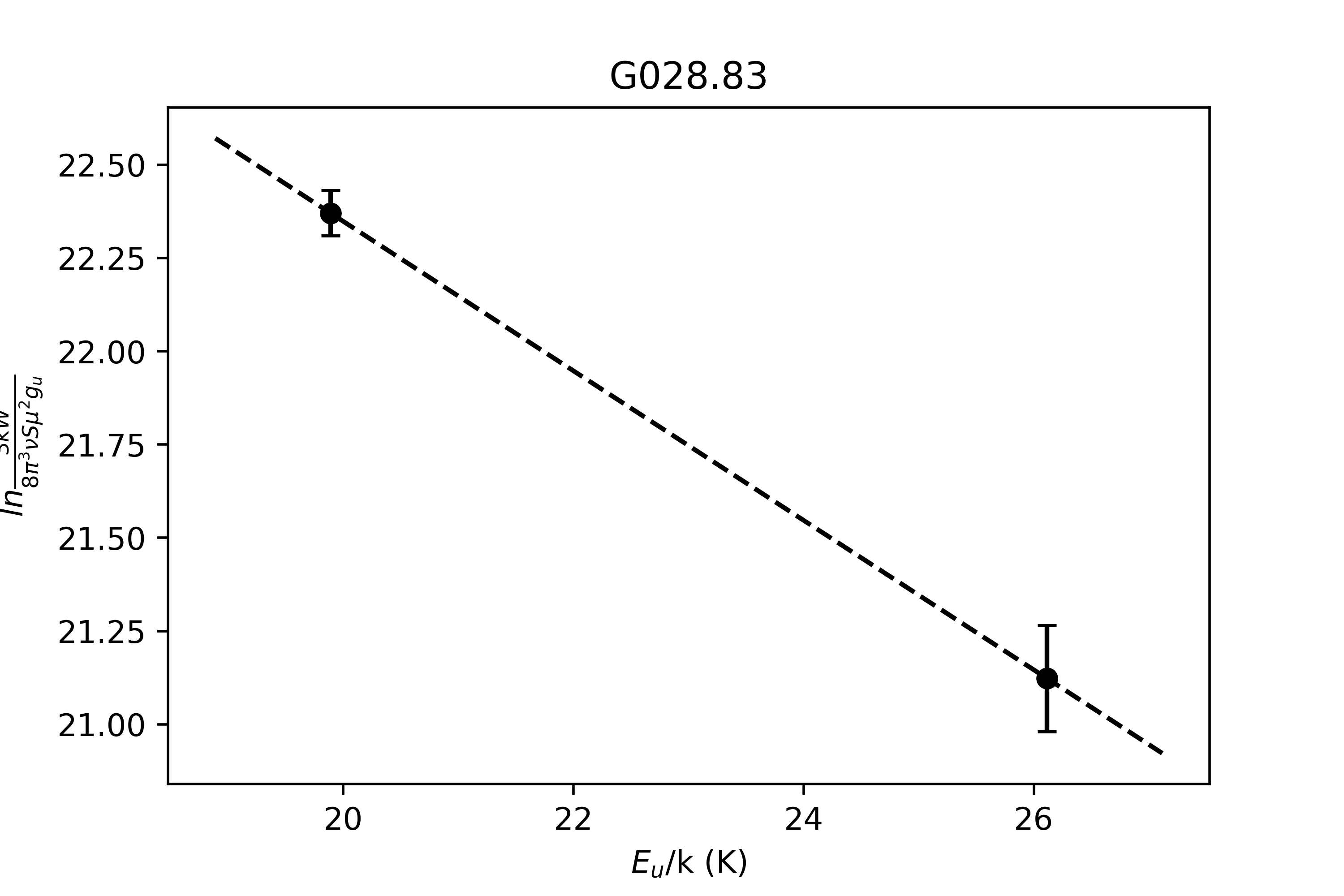}}
	{\includegraphics[width=4.0cm]{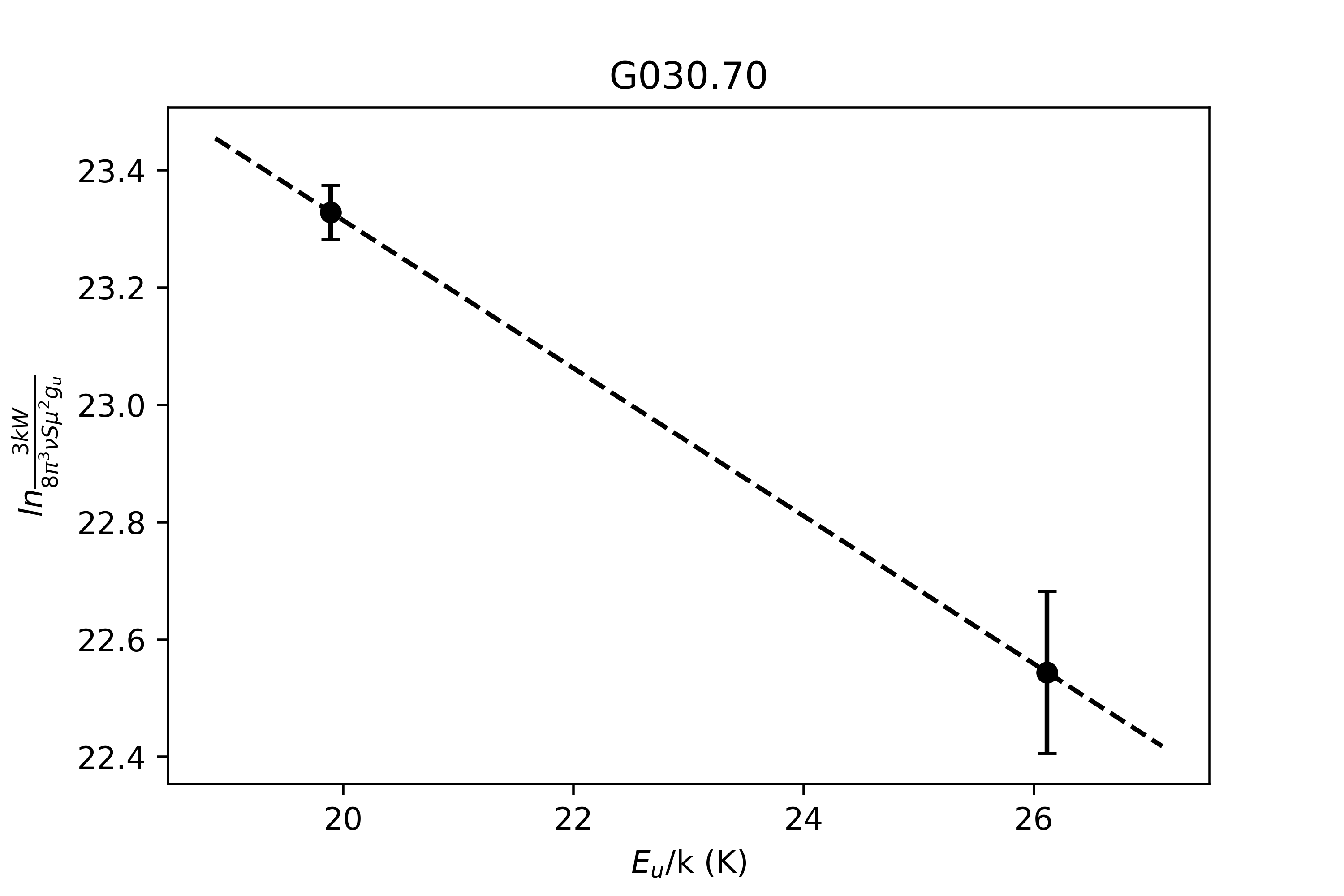}}
	{\includegraphics[width=4.0cm]{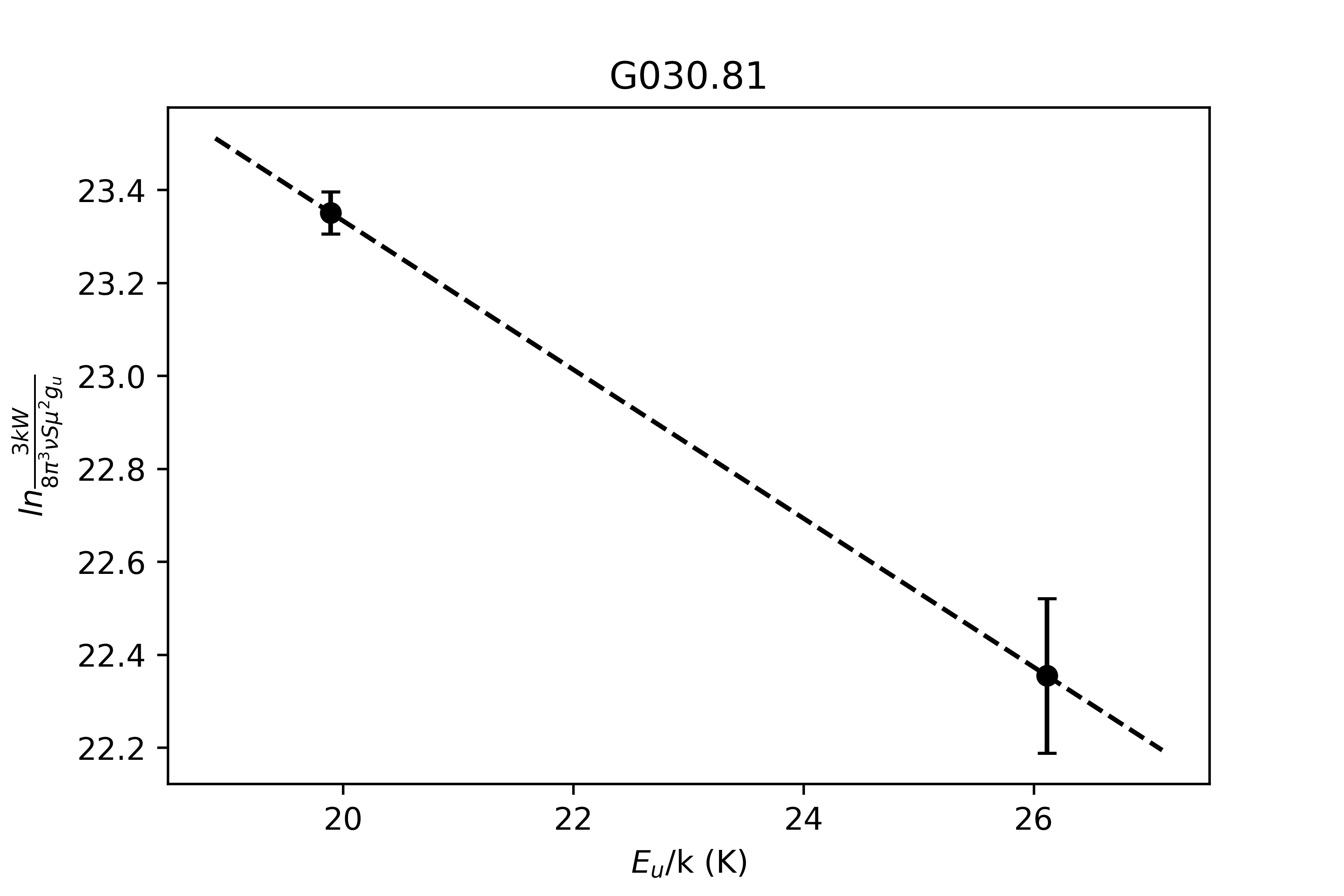}}
	{\includegraphics[width=4.0cm]{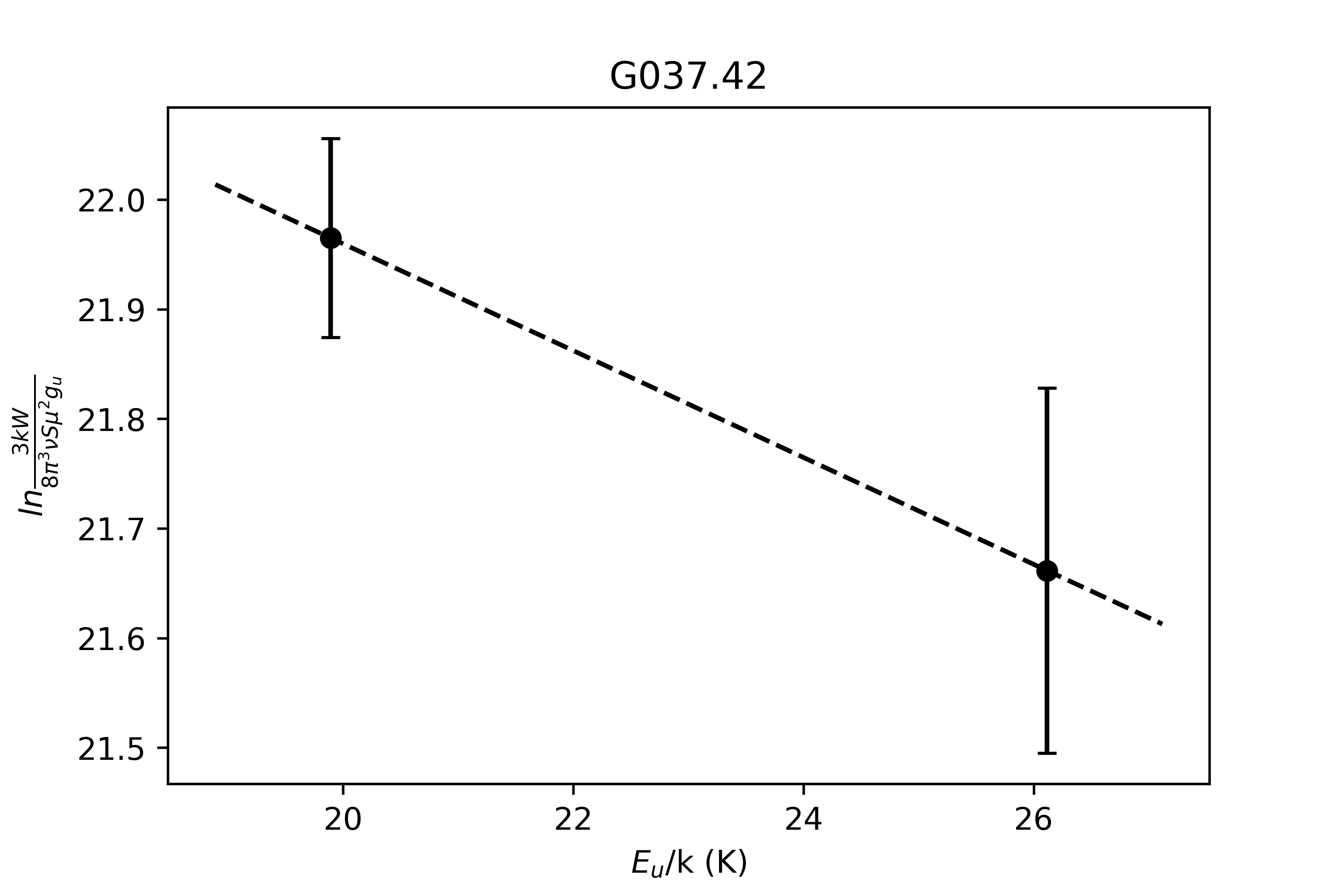}}
	{\includegraphics[width=4.0cm]{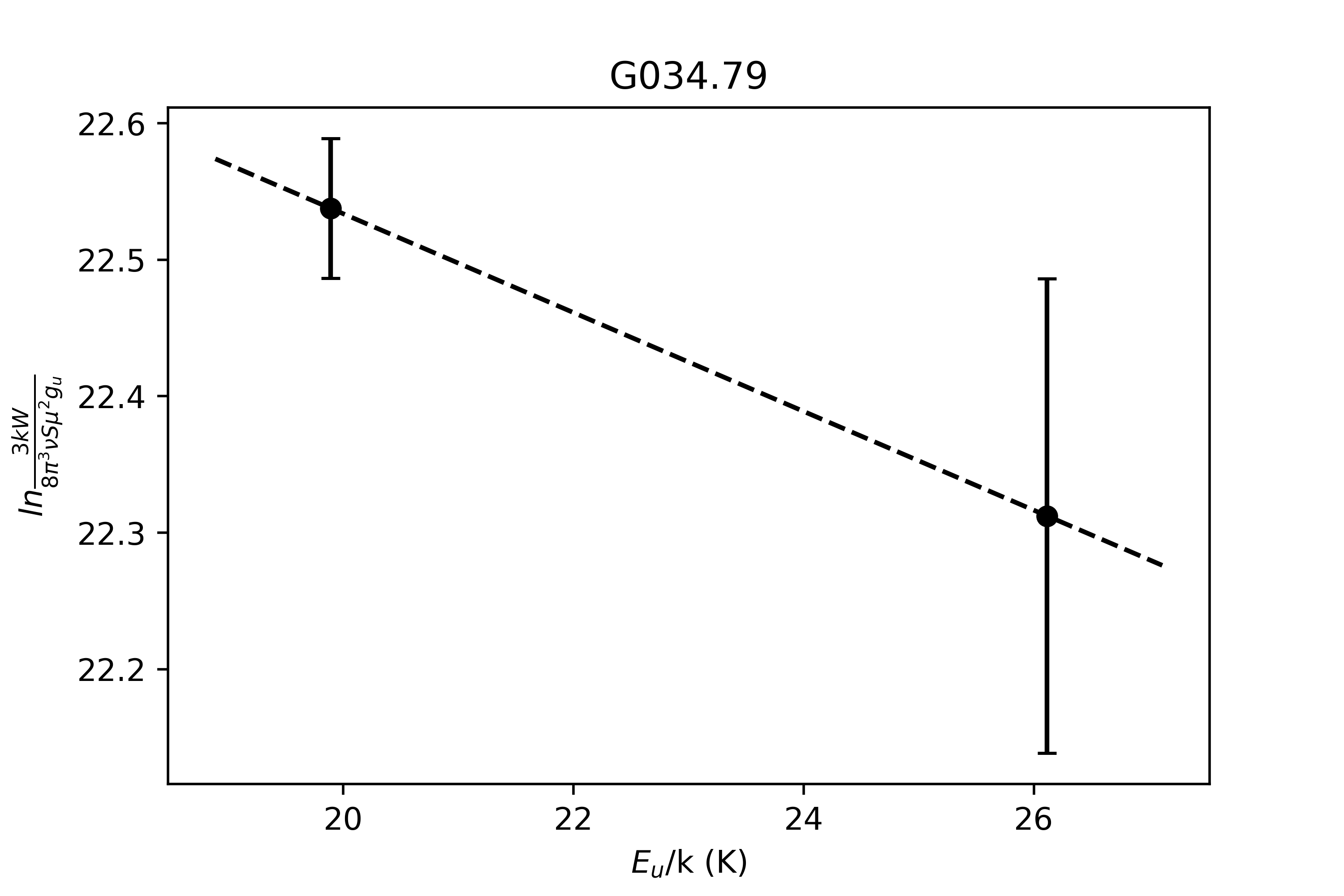}}
	{\includegraphics[width=4.0cm]{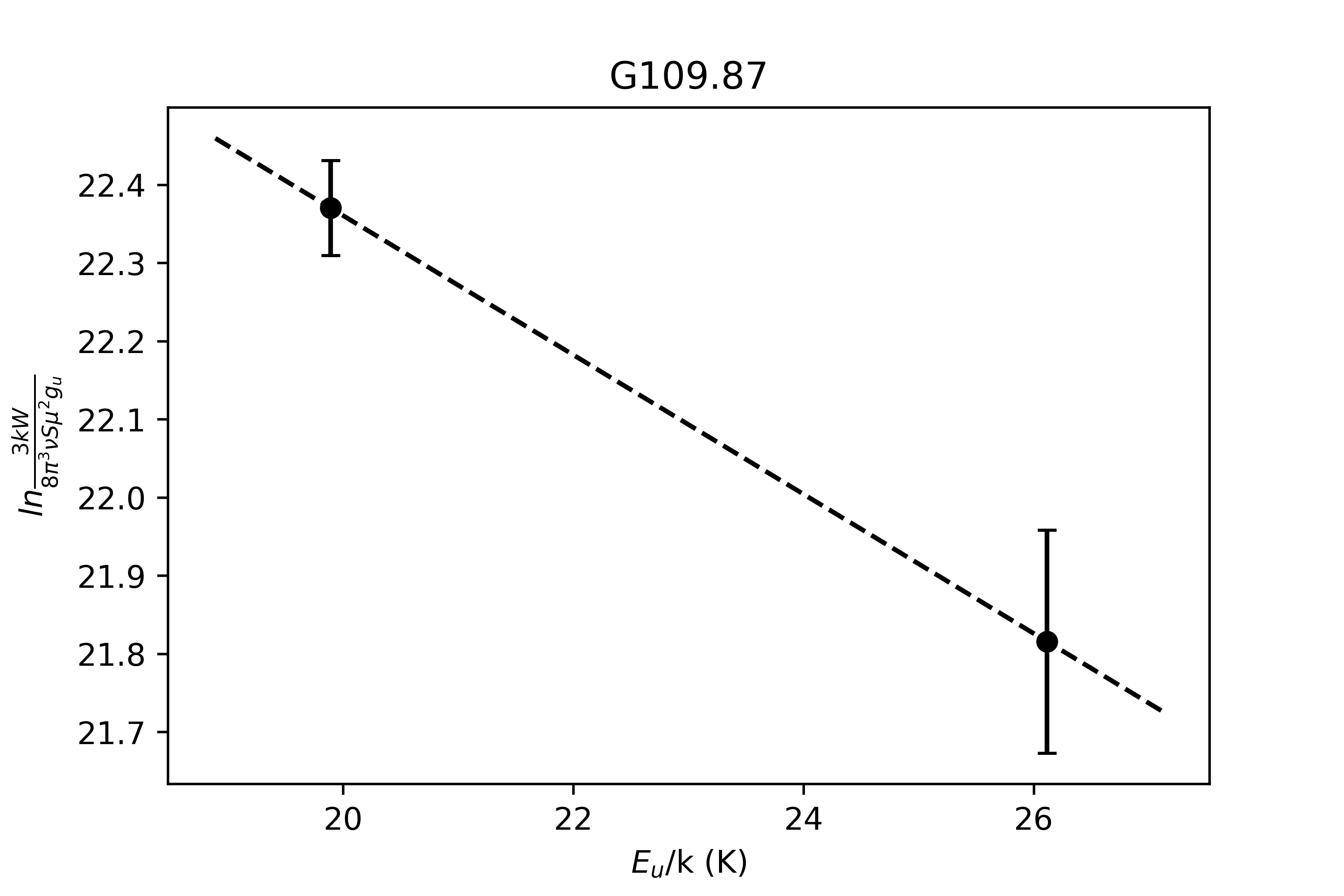}}
	\caption{Rotational diagrams of CCS for sources with  CCS ($J_{N}$ = $8_{7}-7_{6}$) and CCS ($J_{N}$ = $7_{7}-6_{6}$) line detections.}
	\label{fig:Rotational diagram}
\end{figure*}

The CCS molecular transitions are usually assumed to be optically thin \cite[e.g.,][]{2021SCPMA..6479511X}, which can be supported by the low values of the main beam brightness temperature in our sample (\textless0.24 K).
With the assumption of Local Thermodynamic Equilibrium (LTE), optically thin condition and negligible background temperature, the rotational temperature ($T_{rot}$) and column density of CCS can be usually determined by the rotational diagram method with the following formula \citep{1999ApJ...517..209G,2015PASP..127..266M}:
\begin{equation} \label{equ:rotational diagram}  
	   ln\frac{3kW}{8\pi ^{3} \nu S\mu^{2} g_{u}} =  ln\frac{N}{Q(T_{rot})} - \frac{E_{u}}{kT_{rot}},
\end{equation} 
where $W$ is the integrated line intensity of CCS, $T_{rot}$ and $N$ are the rotational temperature and the column density of CCS, respectively. 
Since this method with no background approach will cause an underestimate of $\sim$10-20\% in column density with the Trot of $\sim$ 10 K \cite[Figure 3 in][]{2015PASP..127..266M}, we thus took a 15\% correction to the derived CCS column density.
For 10 UC H{\footnotesize II}s with detections of both CCS lines, the rotational diagrams are shown in Figure \ref{fig:Rotational diagram}. The derived $T_{rot}$ of CCS are listed in column (5) in Table \ref{tab:resultUCHII}, which is consistent with those results derived from the para-NH$_{3}$ (1, 1) and (2, 2) transitions \citep{2010MNRAS.402.2682H,2011ApJ...741..110D,2011MNRAS.418.1689U,2016ApJ...822...59S,2021ApJS..257...39C}, within uncertainties. For 23 UC H{\footnotesize II}s with only CCS $J_{N}$ = $8_{7}-7_{6}$ detection, the column density of CCS was estimated by Equation (\ref{equ:rotational diagram}), with a $T_{rot}$ value of  $\sim$11.93 K (the average $T_{rot}$ of those 10 UC H{\footnotesize II}s with detections of both CCS lines).
The derived column densities of CCS are summarized in Table \ref{tab:resultUCHII}.

\begin{figure*}[htbp]
	\centering  
	{\includegraphics[width=10cm]{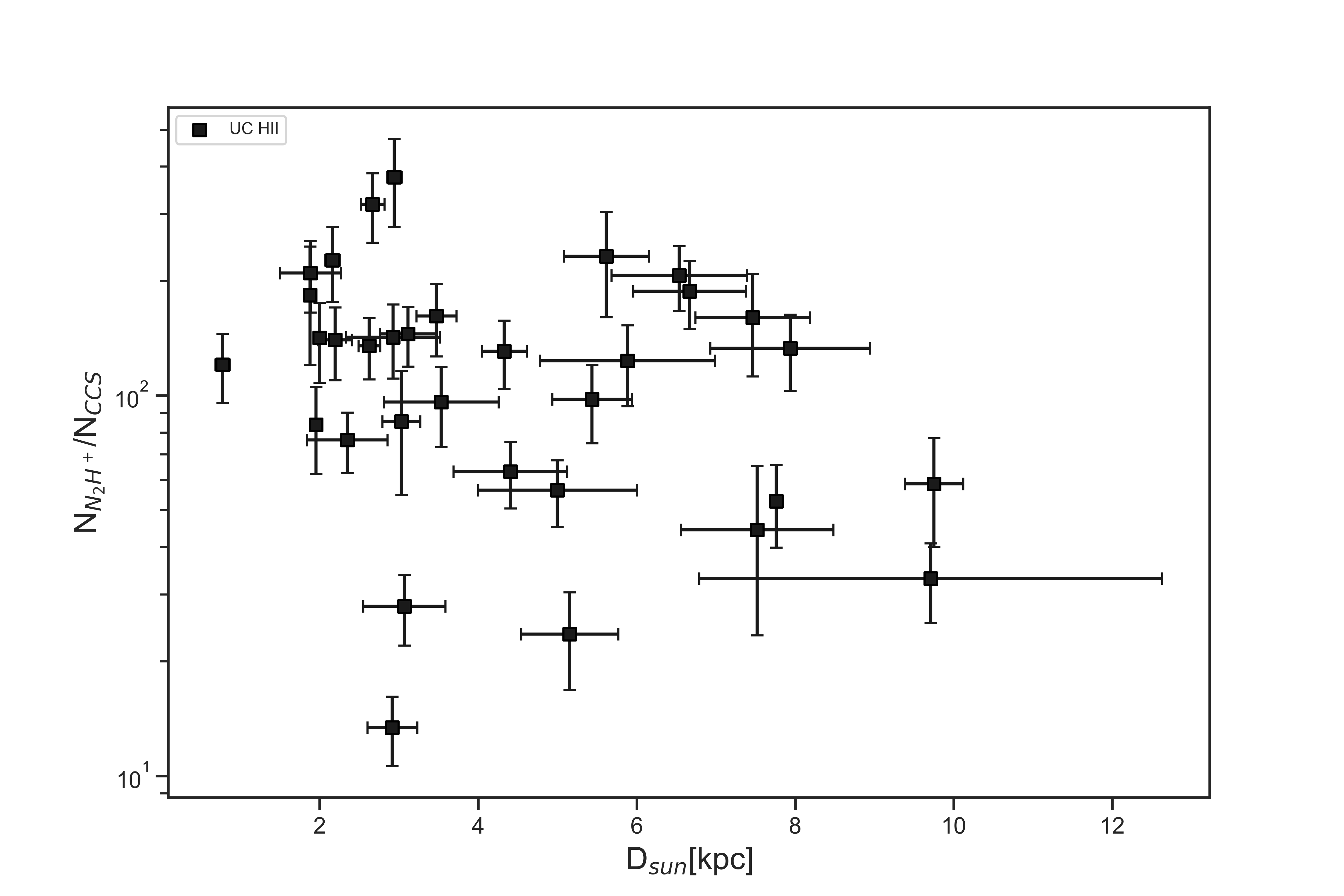}}\hspace{-1cm}
	\caption{$N$(N${_2}$H$^{+}$)/$N$(CCS) against the heliocentric distance; no significant variation can be found between them.
	}
	\label{fig:Distance_N2Hp_CCS}
\end{figure*}

The CCS emission often exhibits a clumpy and spatially different distribution compared to N$_2$H$^+$ \citep{2001ApJ...552..639A,2017ApJS..228...12T}, which may bring uncertainties on our measured $N$(N${_2}$H$^{+}$)/$N$(CCS) ratio. However, both molecules tend to be relative to compact structure, with a spatial distribution size typically of $\leq$10\arcsec \citep[e.g.,][]{2017ApJS..228...12T}, which is less than the beam size of IRAM 30m ($\sim$27\arcsec) we used. Our measured relative intensity of both molecules may not be  affected seriously by their different spatial distributions.
Further the $N$(N${_2}$H$^{+}$)/$N$(CCS) ratio toward our UC H{\footnotesize II} sample was calculated and listed also in Table \ref{tab:resultUCHII}. Our measured results of the $N$(N${_2}$H$^{+}$)/$N$(CCS) ratio may be biased by the beam effect, since the physical scale sampled by the beam varies with distance. Larger distances correspond to a larger physical scale, potentially encompassing more diffuse, low-density gas, impacting the $N$(N${_2}$H$^{+}$)/$N$(CCS) ratio results. To assess this potential bias, we depicted the $N$(N${_2}$H$^{+}$)/$N$(CCS) ratio against heliocentric distance toward our sample. No significant correlation can be found (Figure \ref{fig:Distance_N2Hp_CCS}, which is supported by Pearson test, with a very low correlation coefficient of -0.25. This implies non-significant observational bias on our measured results.

\section{Discussion}\label{sec:Discussion}

\subsection{A good Chemical Evolutionary Indicator, $N$(N${_2}$H$^{+}$)/$N$(CCS)}\label{sec:Evolutionary Indicator of N$_{2}$H$^{+}$/CCS} 
CCS, a well-known carbon-chain molecule, forms from ionic carbon (C$^{+}$) and atomic carbon (C) at the early stage of molecular clouds, making it generally abundant during the initial phases of chemical evolution \citep[e.g.,][]{2003ApJ...593..906A,2007ApJ...663.1174S,2019ApJ...872..154T}. 
On the other hand, the abundance of N${_2}$H$^{+}$ is theoretically expected to increase in the later stages of chemical evolution, as it forms from N${_2}$, whose production is slow in dark clouds \citep[e.g.,][]{1996ApJ...468..761K,2001ApJ...552..639A,2015ApJ...807..120A,2017iace.book.....Y}. Therefore, the ratio of $N$(N${_2}$H$^{+}$)/$N$(CCS) should be served as one indicator for tracing the chemical evolution of HMSFRs.


To check this, i.e., the trend of variation on $N$(N${_2}$H$^{+}$)/$N$(CCS) in HMSFRs stages, we collected observation data of N${_2}$H$^{+}$ and CCS in other evolution stages (HMSC and HMPO), to make comparison on results between them and our UC H{\footnotesize II} sample.  The HMSC data were collected from \cite{2011AA...529L...7F,2023AA...680A..58F} by IRAM 30 m telescope (with a beam size of 26\arcsec) and Chen et al. 2025 (in prep.) by the ARO 12 m telescope (a beam size of 66\arcsec). The HMPO data were taken from Nobeyama 45 m  observation \citep[a beam size of 17\arcsec,][]{2019ApJ...872..154T} and the ARO 12 m observation (a beam size of 66\arcsec, Chen et al. 2025, in prep.). The basic information about the data from Chen et al. (2025, in prep.) are listed in Appendix \ref{sec: AppendixB}.Since the N$_2$H$^+$ and CCS data for each source were obtained using the same telescope—either the IRAM 30\,m, Nobeyama 45\,m, or ARO 12\,m—the beam dilution effect is expected to be similar for both molecules. As our analysis focuses on relative intensity of N$_2$H$^+$ and CCS, the use of archival data from different telescopes with varying beam sizes may bring a non-significant effect on measured ratio results.

\begin{figure*}
	\centering  
	\subfigure[]{\includegraphics[width=5.3cm]{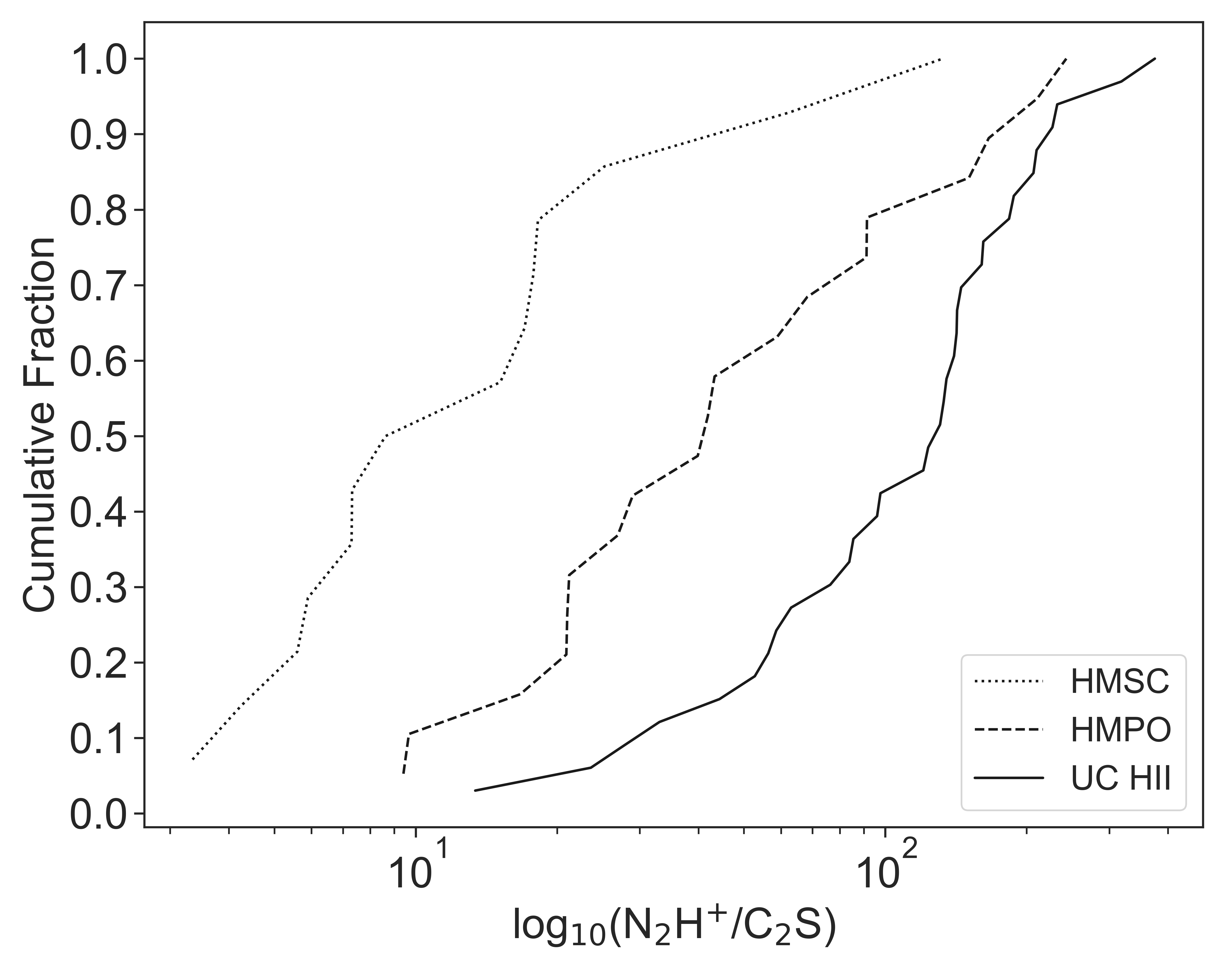}}\hspace{-0.1cm}
	\subfigure[]{\includegraphics[width=5.3cm]{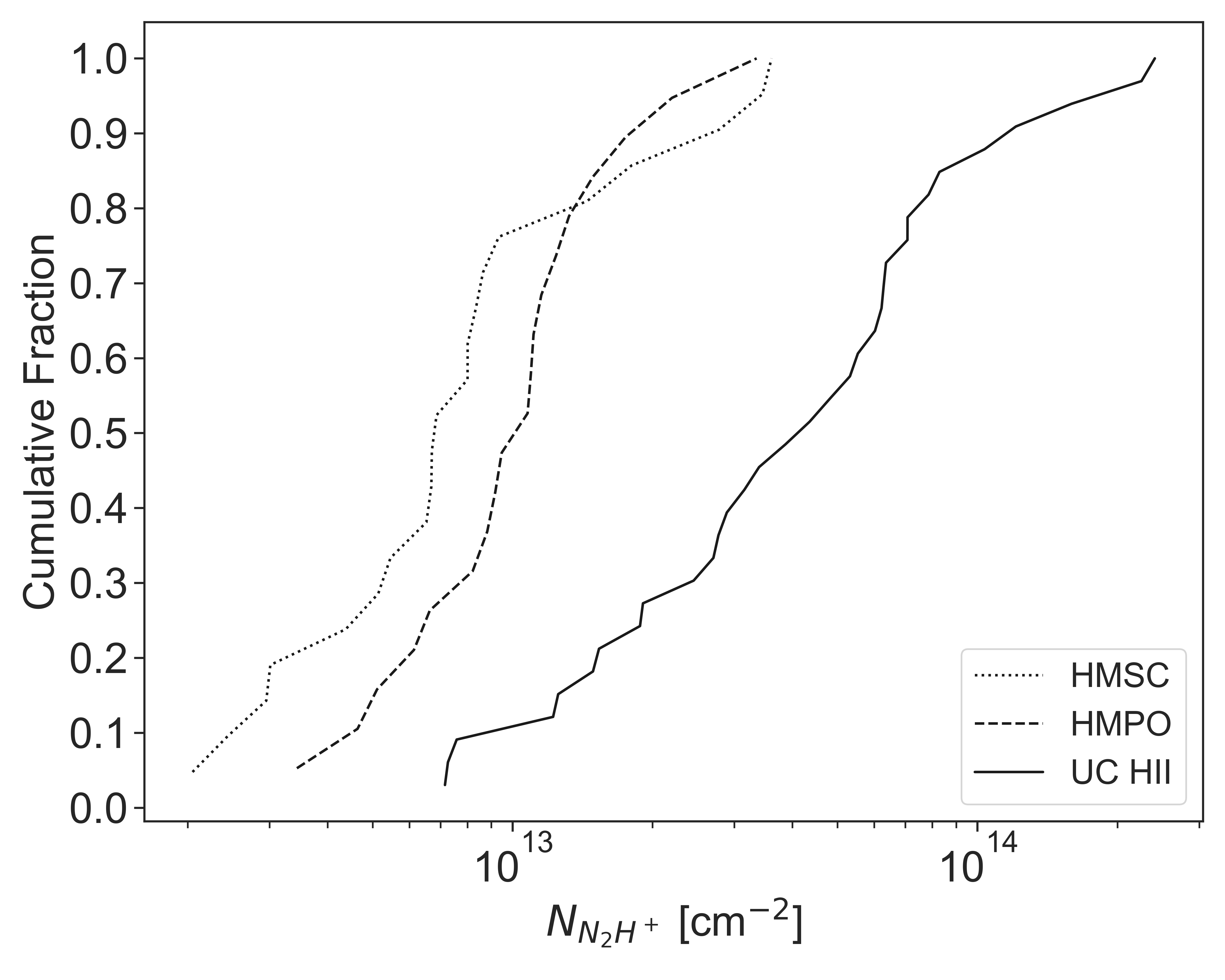}}\hspace{-0.1cm}
	\subfigure[]{\includegraphics[width=5.3cm]{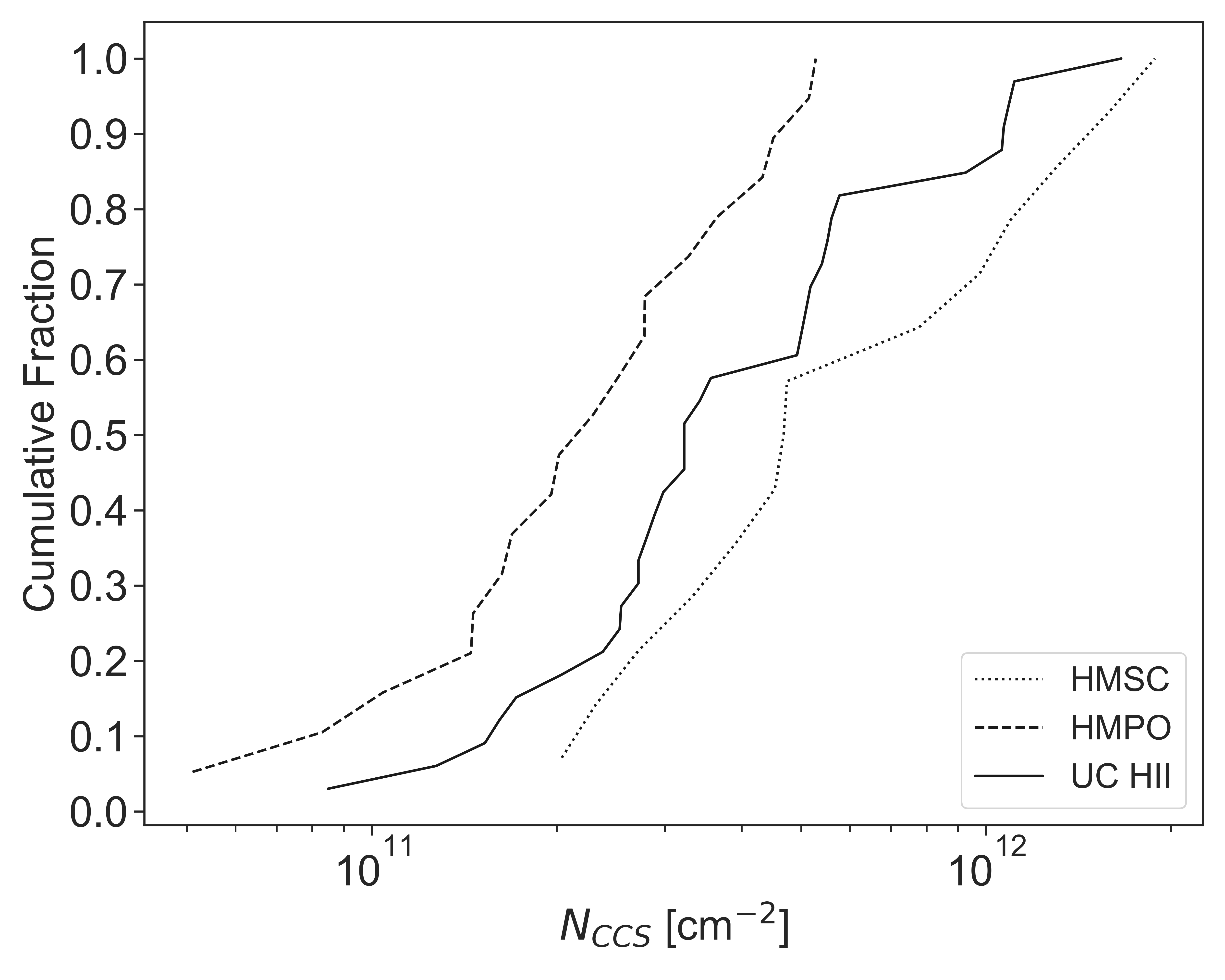}}
	\caption{The cumulative distributions for $N$(N${_2}$H$^{+}$)/$N$(CCS) (a), $N$(N${_2}$H$^{+}$) (b) and $N$(CCS) (c) for HMSCs, HMPOs, and UC H{\footnotesize II}s.
	}
	\label{fig:cumulative distributions}
\end{figure*}

Using those collected N${_2}$H$^{+}$ and CCS data in  HMSCs and HMPOs, we took identical method (see details in Section \ref{sec:Physical parameters}) to determine their physical parameters, including the optical depth, the excitation temperature, the column density of N${_2}$H$^{+}$ and CCS and the ratio of $N$(N${_2}$H$^{+}$)/$N$(CCS) (Table \ref{tab:HMSC_HMPO}). 
The cumulative distribution functions of $N$(N${_2}$H$^{+}$)/$N$(CCS) for both samples and our UC H{\footnotesize II} sample are plotted in Figure \ref{fig:cumulative distributions}(a). Significant difference can be found between those three samples, which is supported by the Kolmogorov-Smirnov (K-S) test statistical results. A chance probability, i.e., those samples from the same parent population, is less than 0.01. This is further supported by the difference on the average $N$(N${_2}$H$^{+}$)/$N$(CCS) values of three samples and corresponding t-test results (see details in Table \ref{tab:mean value}). It shows that the $N$(N${_2}$H$^{+}$)/$N$(CCS) ratio significantly increases from HMSCs to HMPOs and further to UC H{\footnotesize II} regions. 
The increasing $N$(N$_2$H$^+$)/$N$(CCS) ratio may be attributed to the rise in $N$(N$_2$H$^+$) from HMSCs to HMPOs and further to UC H{\footnotesize II} regions, while no significant changes in the CCS column density across those three stages (see detail in Figure \ref{fig:cumulative distributions}b, c; Table \ref{tab:mean value}).

\startlongtable
\renewcommand\tabcolsep{2.0pt} 
		\begin{deluxetable*}{llcccccc}
			\tablecaption{Derived Parameters of CCS and N$_{2}$H$^{+}$ in HMSCs and HMPOs \label{tab:HMSC_HMPO}}
			\tablehead{
				\multirow{3}{*}{Evolutionary} & 
				\multirow{3}{*}{Source} &
				\multicolumn{3}{c}{N$_{2}$H$^{+}$}&
				\colhead{CCS}&
				\multirow{1}{*}{$\frac{\rm N_{2}H^{+}}{\rm CCS}$} &                                  
				\colhead{}   
				\\ \cmidrule(lr){3-5} 
				\colhead{Stage}&
				\colhead{}&
				\colhead{$\tau$}           & 
				\colhead{$T_{ex}$}          & 
				\colhead{$N$}             &  
				\colhead{$N$}             &                                    
				\colhead{} &
				\colhead{References}
				\\  
				\colhead{}&                 
				\colhead{}&                
				\colhead{}& 
				\colhead{(K)}               & 
				\colhead{($\times$ 10$^{13}$ cm$^{-2}$)}           &     
				\colhead{($\times$ 10$^{11}$ cm$^{-2}$)}            &                                  
				\colhead{}                                            &                                  
				\colhead{}                                     
			}
			\decimalcolnumbers
			\startdata
			HMSC & I00117-MM2   & 0.10 $\pm$ 0.10 & 7.00 $\pm$ 1.00  & 0.44 $\pm$ 0.20 & 13.07 $\pm$ 3.46 & 3.35 $\pm$ 1.76    & \cite{2011AA...529L...7F,2023AA...680A..58F} \\
			& AFGL5142-EC  & 0.51 $\pm$ 0.01 & 44.10 $\pm$ 0.10 & 0.69 $\pm$ 0.07 & 4.53 $\pm$ 1.71  & 15.14 $\pm$ 5.92   & \cite{2011AA...529L...7F,2023AA...680A..58F} \\
			& 05358-mm3    & 5.00 $\pm$ 2.00 & 34.00 $\pm$ 4.00 & 3.61 $\pm$ 2.27 & 2.72 $\pm$ 0.84  & 132.55 $\pm$ 92.96 & \cite{2011AA...529L...7F,2023AA...680A..58F} \\
			& G028-C1(MM9) & 3.00 $\pm$ 1.00 & 6.40 $\pm$ 0.40  & 0.93 $\pm$ 0.43 & 15.85 $\pm$ 4.67 & 5.90 $\pm$ 3.21    & \cite{2011AA...529L...7F,2023AA...680A..58F} \\
			& I22134-B     & 2.30 $\pm$ 0.30 & 10.40 $\pm$ 0.40 & 0.24 $\pm$ 0.05 & 3.33 $\pm$ 1.13  & 7.32 $\pm$ 2.89    & \cite{2011AA...529L...7F,2023AA...680A..58F} \\
			& I22134-G     & 3.30 $\pm$ 0.20 & 15.90 $\pm$ 0.30 & 0.51 $\pm$ 0.07 & 2.04 $\pm$ 0.70  & 25.22 $\pm$ 9.30   & \cite{2011AA...529L...7F,2023AA...680A..58F} \\ 
			& 18182-1433-3  & 1.93  $\pm$ 0.76 & 3.01  $\pm$ 0.05 & 0.80  $\pm$ 0.16 & 18.83 $\pm$ 4.96 & 4.25 $\pm$ 1.41    & (Chen et al. 2025, in prep.) \\
			& 18223-1243-3  & 1.04  $\pm$ 0.44 & 5.75  $\pm$ 0.75 & 1.45  $\pm$ 0.26 & 2.32 $\pm$ 0.51  & 62.37 $\pm$ 17.82  & (Chen et al. 2025, in prep.) \\
			& 18247-1147-3  & 1.44  $\pm$ 0.60 & 3.12  $\pm$ 0.08 & 0.80  $\pm$ 0.21 & 10.96 $\pm$ 3.53 & 7.30 $\pm$ 3.02    & (Chen et al. 2025, in prep.) \\
			& 18337-0743-3  & 1.05  $\pm$ 0.48 & 3.81  $\pm$ 0.29 & 0.83  $\pm$ 0.17 & 4.68 $\pm$ 1.17  & 17.81 $\pm$ 5.77   & (Chen et al. 2025, in prep.) \\
			& 18337-0743-7  & 1.15  $\pm$ 0.48 & 3.67  $\pm$ 0.22 & 0.86  $\pm$ 0.20 & 4.74 $\pm$ 1.31  & 18.21 $\pm$ 6.56   & (Chen et al. 2025, in prep.) \\
			& 18385-0512-3  & 1.26  $\pm$ 0.53 & 3.41  $\pm$ 0.15 & 0.67  $\pm$ 0.14 & 7.77 $\pm$ 2.55  & 8.63 $\pm$ 3.35    & (Chen et al. 2025, in prep.) \\
			& 18530+0215-2  & 1.11  $\pm$ 0.47 & 3.83  $\pm$ 0.27 & 0.67  $\pm$ 0.16 & 3.92 $\pm$ 1.24  & 17.05 $\pm$ 6.75   & (Chen et al. 2025, in prep.) \\
			& 19175+1357-4e & 1.05  $\pm$ 0.55 & 3.23  $\pm$ 0.15 & 0.55  $\pm$ 0.15 & 9.76 $\pm$ 3.85  & 5.60 $\pm$ 2.71    & (Chen et al. 2025, in prep.) \\
			HMPO & 05358+3543 & 1.60  $\pm$ 0.50 & 5.50  $\pm$ 0.40 & 1.49  $\pm$ 0.35 & 1.63  $\pm$ 0.54 & 91.43 $\pm$ 36.81   & \cite{2019ApJ...872..154T} \\
			& 05553+1631 & 1.40  $\pm$ 1.30 & 3.00  $\pm$ 0.10 & 0.51  $\pm$ 0.35 & 5.28  $\pm$ 2.81 & 9.67 $\pm$ 8.45     & \cite{2019ApJ...872..154T} \\
			& 19220+1432 & 0.80  $\pm$ 0.50 & 4.70  $\pm$ 0.80 & 0.82  $\pm$ 0.26 & 1.96  $\pm$ 0.81 & 41.9 $\pm$ 21.87    & \cite{2019ApJ...872..154T} \\
			& 19410+2336 & 1.90  $\pm$ 0.60 & 8.40  $\pm$ 0.80 & 3.35  $\pm$ 0.75 & 2.02  $\pm$ 0.42 & 166.11 $\pm$ 50.92  & \cite{2019ApJ...872..154T} \\
			& 19413+2332 & 2.90  $\pm$ 1.40 & 3.00  $\pm$ 0.10 & 1.08  $\pm$ 0.49 & 5.15  $\pm$ 3.18 & 20.93 $\pm$ 16.04   & \cite{2019ApJ...872..154T} \\
			& 20051+3435 & 2.20  $\pm$ 1.20 & 3.00  $\pm$ 0.10 & 0.95  $\pm$ 0.44 & 4.51  $\pm$ 2.83 & 21.03 $\pm$ 16.45   & \cite{2019ApJ...872..154T} \\
			& 20126+4104 & 1.10  $\pm$ 0.50 & 8.80  $\pm$ 1.40 & 2.20  $\pm$ 0.48 & 1.46  $\pm$ 0.46 & 150.73 $\pm$ 57.54  & \cite{2019ApJ...872..154T} \\
			& 20332+4124 & 1.60  $\pm$ 0.70 & 3.50  $\pm$ 0.20 & 0.92  $\pm$ 0.31 & 4.32  $\pm$ 2.21 & 21.23 $\pm$ 12.97   & \cite{2019ApJ...872..154T} \\
			& 20343+4129 & 2.00  $\pm$ 0.60 & 4.40  $\pm$ 0.20 & 1.24  $\pm$ 0.32 & 0.51  $\pm$ 0.31 & 242.98 $\pm$ 161.78 & \cite{2019ApJ...872..154T} \\
			& 22134+5834 & 0.80  $\pm$ 0.70 & 3.60  $\pm$ 0.50 & 0.34  $\pm$ 0.17 & 3.65  $\pm$ 2.47 & 9.42 $\pm$ 7.91     & \cite{2019ApJ...872..154T} \\
			& 23033+5951 & 0.80  $\pm$ 0.50 & 6.30  $\pm$ 1.40 & 1.11  $\pm$ 0.27 & 2.78  $\pm$ 1.03 & 39.87 $\pm$ 17.59   & \cite{2019ApJ...872..154T} \\
			& 23139+5939 & 1.30  $\pm$ 0.60 & 4.00  $\pm$ 0.30 & 0.88  $\pm$ 0.26 & 3.28  $\pm$ 1.78 & 26.93 $\pm$ 16.58   & \cite{2019ApJ...872..154T}\\
			& 18151-1208    & 0.67  $\pm$ 0.43 & 5.19  $\pm$ 1.07 & 0.66  $\pm$ 0.15 & 2.29 $\pm$ 0.58  & 28.99 $\pm$ 9.74   & (Chen et al. 2025, in prep.) \\
			& 18223-1243    & 1.07  $\pm$ 0.44 & 5.41  $\pm$ 0.64 & 1.10  $\pm$ 0.14 & 2.53 $\pm$ 0.62  & 43.31 $\pm$ 11.96  & (Chen et al. 2025, in prep.) \\
			& 18264-1152    & 0.53  $\pm$ 0.39 & 8.31  $\pm$ 3.05 & 1.32  $\pm$ 0.21 & 1.45 $\pm$ 0.25  & 91.22 $\pm$ 21.44  & (Chen et al. 2025, in prep.) \\
			& 18308-0841    & 0.58  $\pm$ 0.41 & 9.07  $\pm$ 3.21 & 1.75  $\pm$ 0.22 & 0.83 $\pm$ 0.26  & 210.78 $\pm$ 71.15 & (Chen et al. 2025, in prep.) \\
			& 18454-0136    & 0.68  $\pm$ 0.44 & 3.82  $\pm$ 0.49 & 0.46  $\pm$ 0.15 & 2.78 $\pm$ 0.97  & 16.70 $\pm$ 7.85   & (Chen et al. 2025, in prep.) \\
			& 18488+0000    & 0.49  $\pm$ 0.42 & 5.65  $\pm$ 1.82 & 0.61  $\pm$ 0.12 & 1.04 $\pm$ 0.33  & 58.89 $\pm$ 21.77  & (Chen et al. 2025, in prep.) \\
			& 18530+0215    & 0.63  $\pm$ 0.40 & 7.06  $\pm$ 1.88 & 1.15  $\pm$ 0.20 & 1.69 $\pm$ 0.33  & 68.20 $\pm$ 17.55  & (Chen et al. 2025, in prep.)
			\enddata
			\tablecomments{
				Column(1): evolutionary stage; column(2): source name; column(3)-(5): the optical depth, the excitation temperature, and the column density of N$_{2}$H$^{+}$; column(5): the column density of CCS; column(7): ratio of $N$(N${_2}$H$^{+}$)/$N$(CCS); column(8): references.
			}
		\end{deluxetable*}

\begin{figure*}[htbp]
	\centering  
	\begin{minipage}[c]{1\linewidth}
		\centering
		\includegraphics[width=0.85\linewidth]{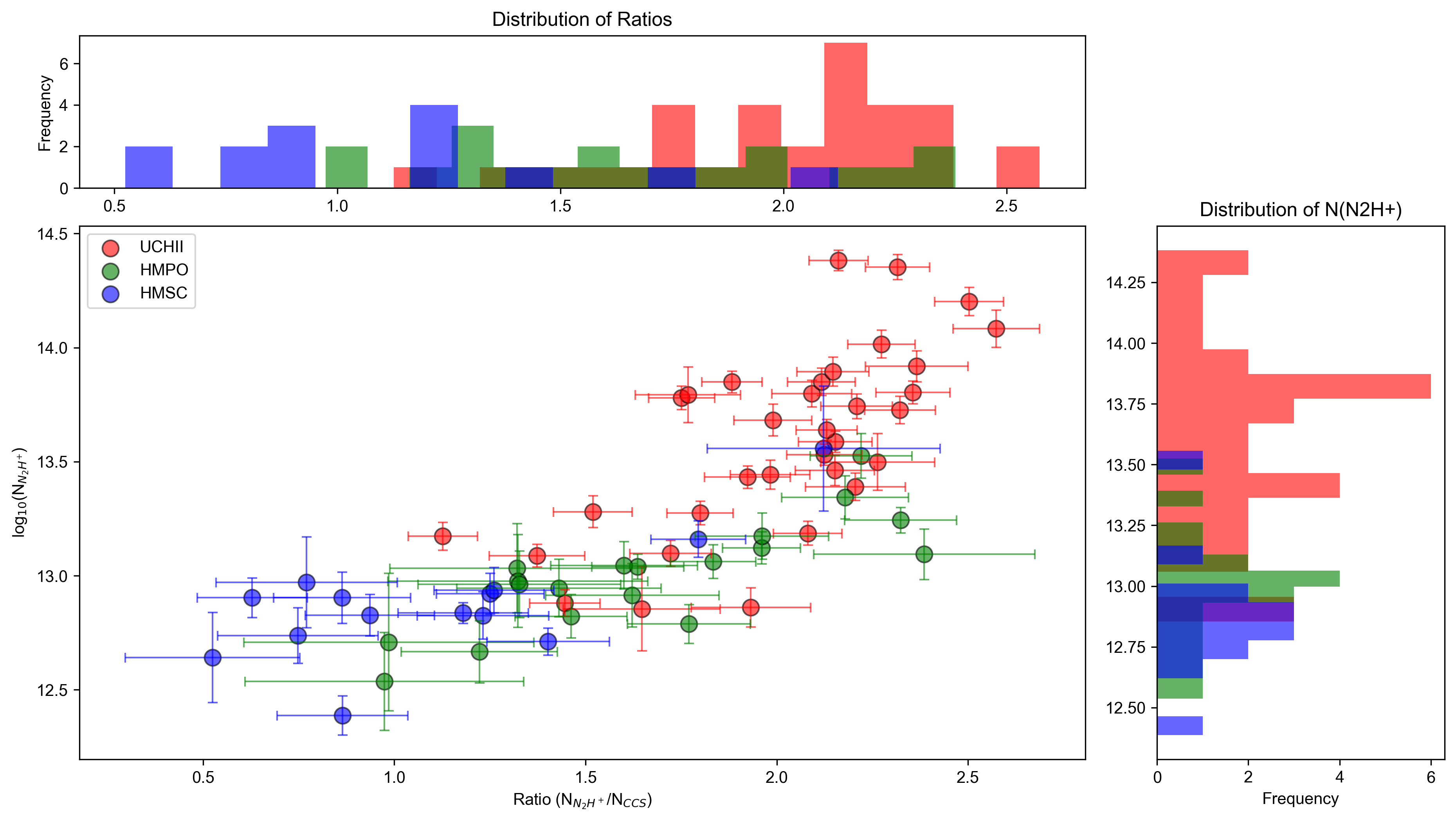}
	\end{minipage}
	\caption{The $N$(N${_2}$H$^{+}$)/$N$(CCS) ratio is plotted against the $N$(N${_2}$H$^{+}$) and the  histograms of the x-axis and y-axis on the top and right panel for the three groups (HMSC, HMPO, UC H{\footnotesize II}) are showed. Sources at three stages can be divided distinctly and a strong correlation between the ratio and the $N$(N${_2}$H$^{+}$) can be found. 
	}
	\label{fig:ratio_N2Hp_CCS}
\end{figure*}

\renewcommand\tabcolsep{25.0pt} 
\begin{deluxetable*}{lc}
	\tablecaption{Average values of the $N$(N${_2}$H$^{+}$)/$N$(CCS) for samples at different stage, and statistical test results between different samples. \label{tab:mean value}}
	\tablehead{
		\colhead{Evolutionary Stage} &  \colhead{$N$(N${_2}$H$^{+}$)/$N$(CCS)}  \\
		\colhead{} & \colhead{}
	}
	\startdata
	HMSC                                            & 23.62 $\pm$ 6.87  \\
	HMPO                                           & 71.60 $\pm$ 10.78 \\
	UC H{\footnotesize II} region  & 129.93 $\pm$ 6.47 \\
	K-S test p-value$^{a}$              & \textless{}0.01  \\
	K-S test p-value$^{b}$             & \textless{}0.01             \\
	K-S test p-value$^{c}$          & \textless{}0.01  \\
	t-test p-value$^{a}$                  & 0.03             \\
	t-test p-value$^{b}$                            & 0.01            \\
	t-test p-value$^{c}$                    & \textless{}0.01 
	\enddata
	\tablecomments{K-S test and t-test chance probabilities for comparison between: $^{a}$HMSCs and HMPOs, $^{b}$ HMPOs and UC H{\footnotesize II}s, and $^{c}$HMSCs and UC H{\footnotesize II}s.}
\end{deluxetable*}

To better illustrate the trend of the $N$(N${_2}$H$^{+}$)/$N$(CCS) ratio, we presented a plot of $N$(N${_2}$H$^{+}$) against the $N$(N${_2}$H$^{+}$)/$N$(CCS) ratio in Figure \ref{fig:ratio_N2Hp_CCS}. The plot clearly shows that sources at those three stages can be divided distinctly, and a strong correlation, i.e., an increasing $N$(N${_2}$H$^{+}$)/$N$(CCS) ratio from HMSCs to HMPOs, and further to UC H{\footnotesize II}s. This suggests that the $N$(N${_2}$H$^{+}$)/$N$(CCS) ratio can serve as a reliable chemical evolutionary indicator in HMSFRs.

\subsection{Comparison with the chemical modeling}\label{sec: Comparison with the chemical modeling} 

In Section \ref{sec:Evolutionary Indicator of N$_{2}$H$^{+}$/CCS}, our measured results show that the $N$(N$_{2}$H$^{+}$)/$N$(CCS) ratio increases obviously across HMSCs, HMPOs, and UC H{\footnotesize II}s, i.e., the ratio can be taken as one good indicator of HMSFRs. To understand better this clock indicator and enhance our comprehension of the high-mass star formation process, we utilized a gas-grain chemical model \cite[GGCHEMPY \footnote{GGCHEMPY is available on \url{https://github.com/JixingGE/GGCHEMPY}} from][]{2022RAA....22a5004G} to constrain physical parameters and chemical ages from the initial HMSCs phase to the later stages of UC H{\footnotesize II}s. This gas-grain chemical model was developed on the basis of the gas-grain chemical processes described in the literature \cite[e.g.,][]{1992ApJS...82..167H,2010A&A...522A..42S}. This model integrates gas-grain chemical processes, encompassing both gas-phase reactions and dust surface reactions. The interconnection arises through accretion and desorption processes of neutral species \citep[e.g.,][]{1992ApJS...82..167H,2010A&A...522A..42S}. The gas-grain reaction network\footnote{The network is sourced from the KIDA database: \url{http://kida.astrophy.ubordeaux.fr/networks.html}.} initially presented by \cite{2010A&A...522A..42S} is employed and updated. The model includes reactive desorption \citep{2007A&A...467.1103G}, as well as CO and H$_{2}$ self-shielding \citep{1996A&A...311..690L}, alongside thermal and cosmic-ray-induced desorption processes \citep{1993MNRAS.261...83H}. 




We adopted one simplified, spatially uniform physical model \citep{2015A&A...578A..55S} to investigate the chemical evolution under typical physical environment  representative of different evolutionary stages, rather than to model the spatial distribution of physical properties in detail. The gas and dust temperatures are assumed to be equal (i.e., $T_{\rm gas} = T_{\rm dust}$). To explore the chemical evolution across different physical environments, we constructed a two-dimensional parameter grid of temperature and density. For each evolutionary stage, the temperature was uniformly sampled with 200 values across the following ranges: 15–35~K for HMSCs, 50–250~K for HMPOs, and 100–500~K for UC~H{\footnotesize II}s \citep{2014A&A...563A..97G}. The gas density ($\rho$) was sampled with 50 logarithmically spaced values from $1.0 \times 10^{5}$ to $5.0 \times 10^{9}$~cm$^{-3}$. Each $(T, \rho$) pair defines a unique static physical condition under which the chemical evolution was simulated from 10$^{4}$ to 10$^{6}$ yr. The cosmic-ray particle (CRP) ionization rate ($\zeta_{\rm CR}$), visual extinction ($A_{v}$), and gas-to-dust mass ratio ($R$) were fixed at $5 \times 10^{-17}$~s$^{-1}$, 10~mag, and 100, respectively. And the initial abundances \citep{2014A&A...563A..97G} used for simulating the chemistry HMSC stage are listed in Table \ref{tab:initial abundance}.

\begin{deluxetable*}{lc}[htbp]
	\tablecaption{Initial atomic and molecular abundances \label{tab:initial abundance}}
	\tablehead{
		\colhead{Species} & \colhead{Relative abundance} 
	}
	\startdata
	H${_2}$ & 0.499 \\
	H & 2.00E-3  \\
	He & 9.75E-2 \\
	C & 7.86E-5  \\
	N & 2.47E-5  \\
	O & 1.80E-4  \\
	S & 8.00E-7  \\
	Si & 3.00E-9  \\
	Na & 2.25E-9  \\
	Mg & 1.09E-8  \\
	Fe & 2.74E-9  \\
	P & 2.16E-10  \\
	Cl & 1.00E-9  \\
	\enddata
\end{deluxetable*}



The modeled abundances of N$_2$H$^+$ and CCS under different temperature and density conditions can be obtained. And then we converted the modeled abundances into column density assuming a line-of-sight thickness of 1 pc (see detail in Appendix \ref{sec: Model column density calculation}). To obtain the best-fit chemical model at the best-fit timescale, we adopted the confidence criterion to evaluate modeling results \citep[e.g.,][]{2007A&A...467.1103G,2008ApJ...681.1385H,2011ApJ...743..182H,2019A&A...622A.185W,2021A&A...648A..72W,2025arXiv250218819W}. For each source ($i$) with detections of both lines, the agreement at each time step between the modeled column density $N_{i(t)}$ and the observed column density $N_{\text{obs},{i(t)}}$ is quantified by the confidence level $\kappa_{i(t)}$, defined as:
\begin{equation}\label{equ:confidence criterion}
	\kappa_{i(t)}=erfc\left(\frac{|\log\left(N_{i(t)}\right)-\log\left(N_{\text{obs},i(t)}\right)|}{\sqrt{2}\sigma}\right).
\end{equation}
where erfc is the complementary error function (erfc = 1 - erf), and the standard deviation is set to $\sigma = 1$. A higher value of $\kappa_{i(t)}$ (ranging from 0 to 1) indicates better agreement between the model and observation while $\kappa_{i(t)} = 0.317$ corresponds to a deviation of one order of magnitude. We calculated the sum of the $\kappa_{(t)}$ at each time step for each model. An iterative search was then performed to find the maximum total $\kappa_{(t)}$ value, representing the best-fit evolutionary time under the best-fit model.
For modeling chemistry in stages beyond HMSCs, we utilized the chemical abundances derived from the best-fit model of the preceding evolutionary stage as the initial values. 
This enabled us to model an approximately steady warming of the matter throughout the evolution of the high-mass star-forming clouds \citep{2014A&A...563A..97G}. The best-fit model and the best-fit chemical timescale for these three evolutionary stage, including the corresponding density, temparature, are listed in Table \ref{tab: best-fit model}.

\begin{figure*}[htbp]
	\centering  
	\subfigure[]{\includegraphics[width=5.3cm]{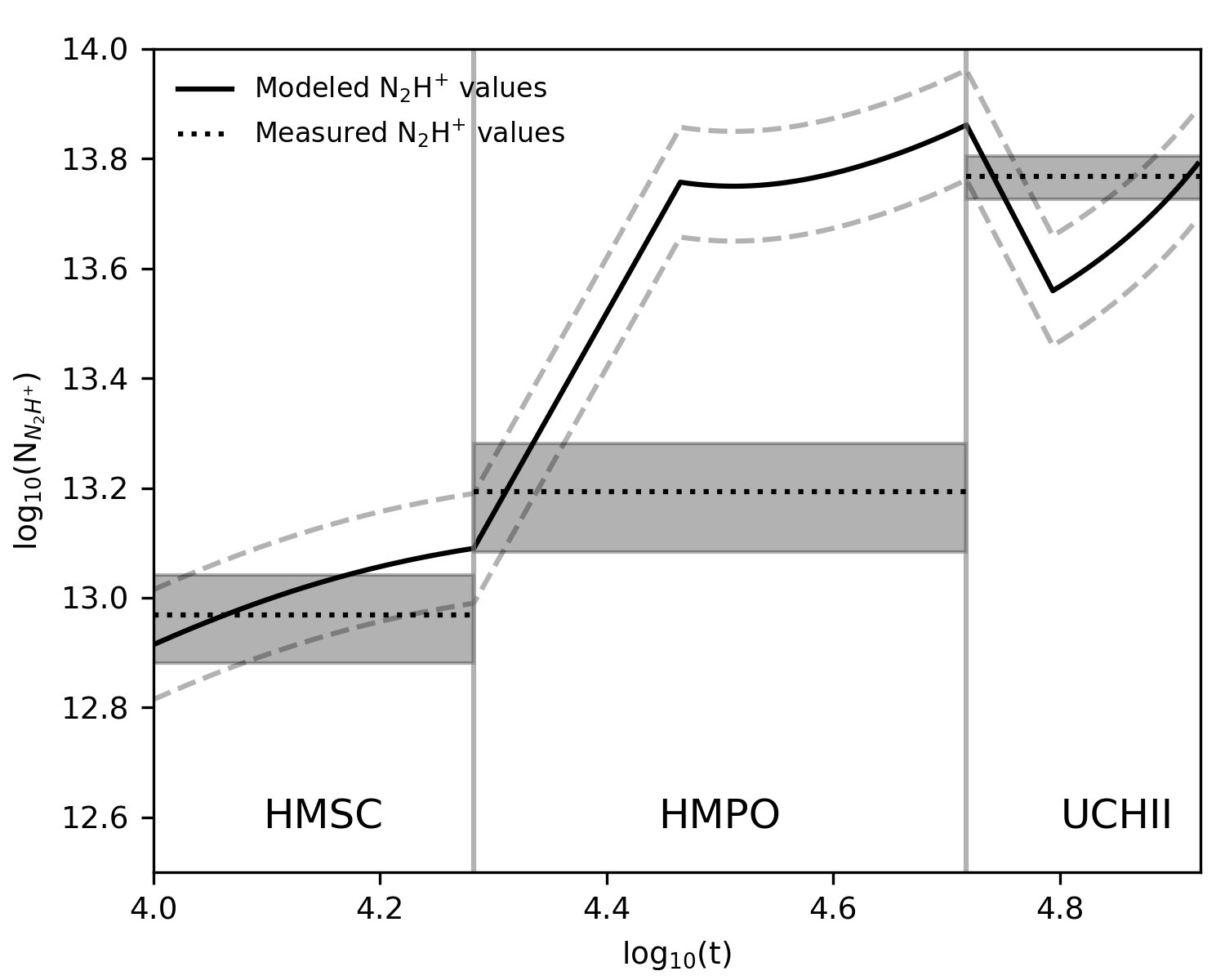}}\hspace{-0.1cm}
	\subfigure[]{\includegraphics[width=5.3cm]{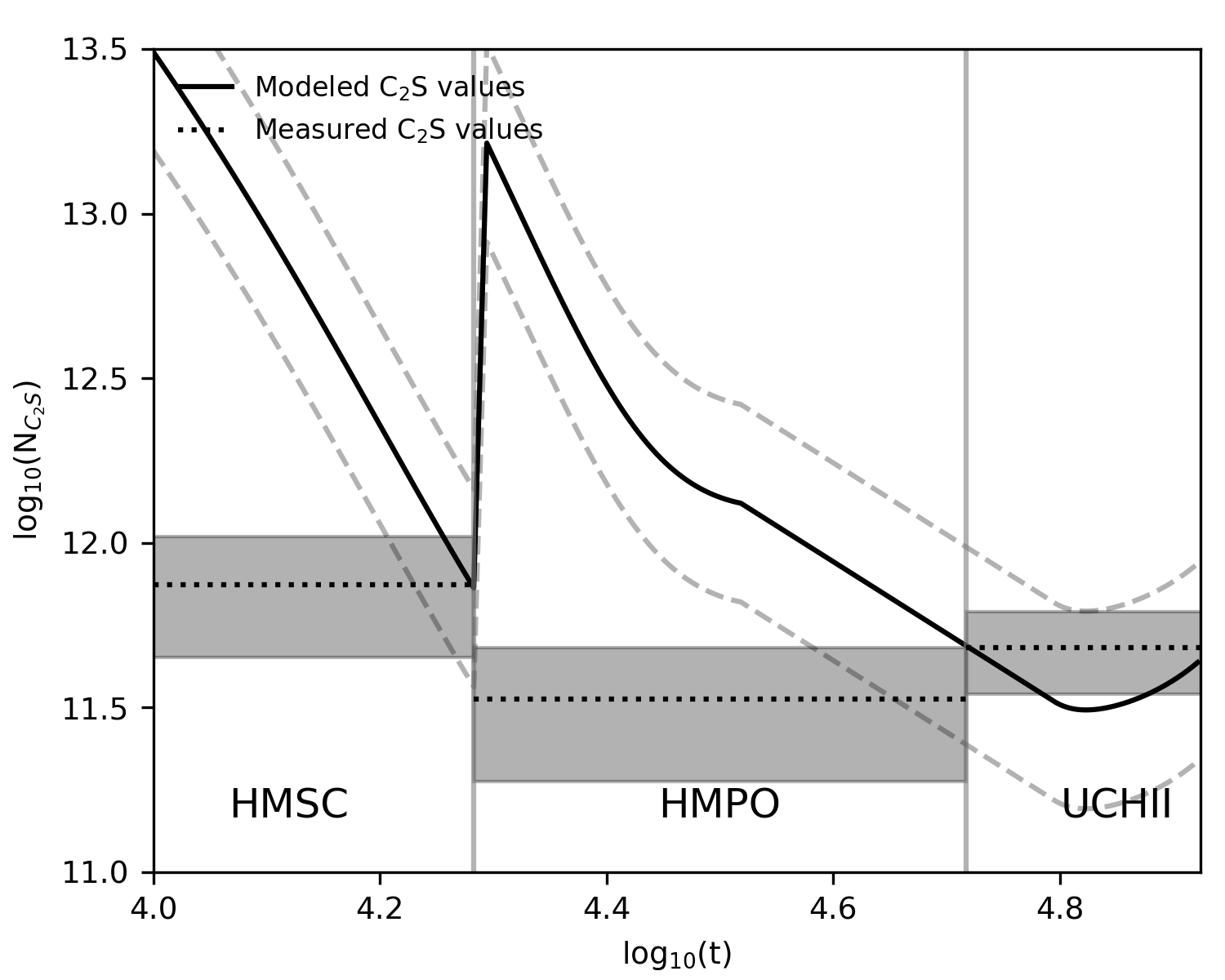}}\hspace{-0.1cm}
	\subfigure[]{\includegraphics[width=5.3cm]{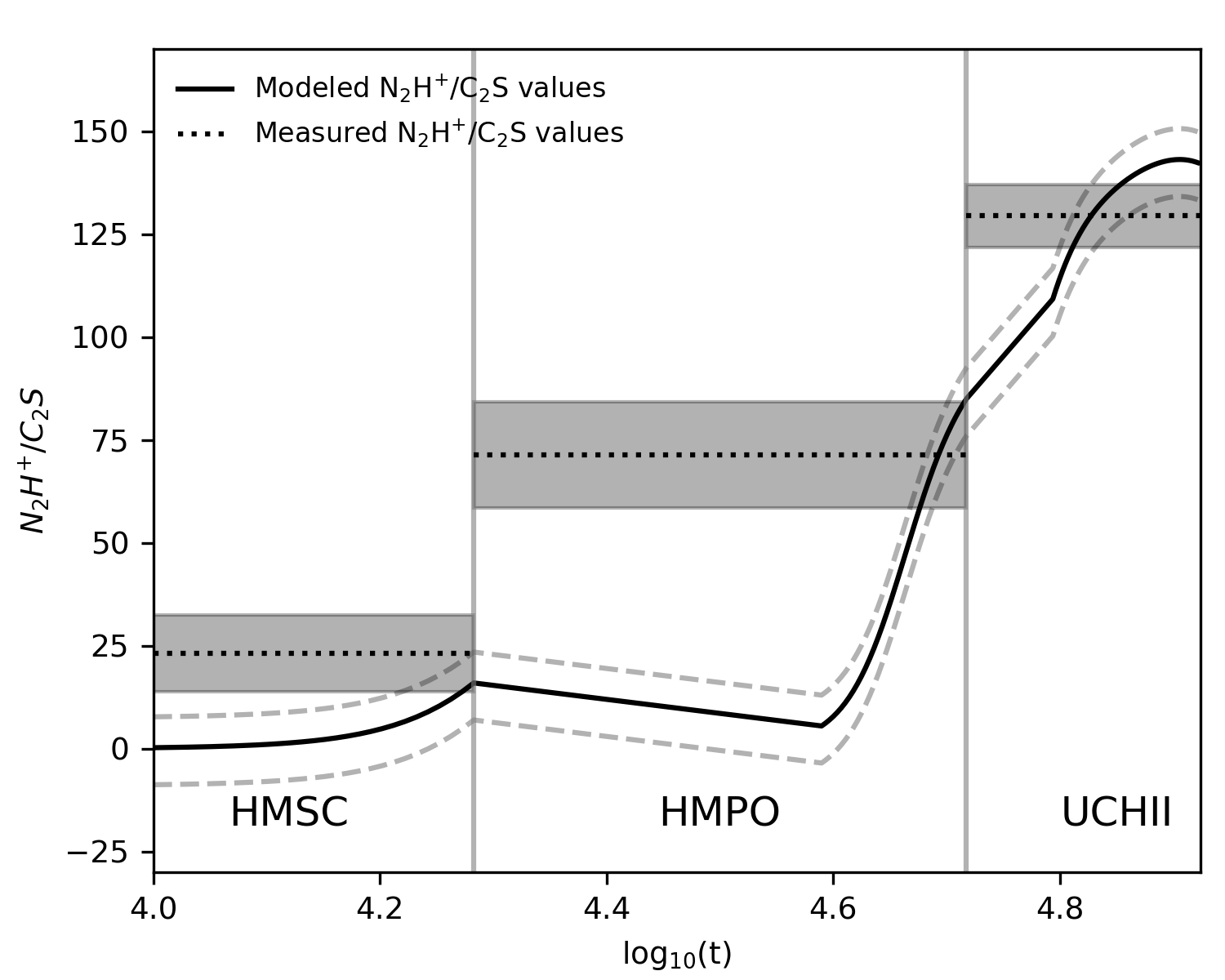}}
	\caption{
	The modeled column density for N$_{2}$H$^{+}$ (a) and CCS (b), and their ratio (c) from our best-fit models were plotted as a function of the age for HMSC, HMPO and UC H{\footnotesize II} stages (solid lines with dashed lines for standard deviation). Those dotted lines show the average values of measured $N$(N$_{2}$H$^{+}$), $N$(CCS) and their ratio at different stages, with the area for the confidence levels. 
	}
	\label{fig:Model_density}
\end{figure*}

The temperature of these three best-fit model tends to increase from HMSCs to HMPOs futher to UC H{\footnotesize II}s, while the best-fit timescale are 19 179, 32 990 and 34 286 yr, respectively (Table \ref{tab: best-fit model}). 
In total, our best-fit timescale of the whole process is 86 455 yr, which is consistent with  the typical high mass star formation age of $\sim$10$^{5}$ yr \citep[e.g.,][]{2003ApJ...585..850M,2014prpl.conf..149T}. And it is also comparable to previous modeling results, e.g.,  $\sim$ 125 000 yr from \cite{2014A&A...563A..97G} and $\sim$ 85 000 yr from \cite{2015A&A...579A..80G}.
The modeled column densities of N$_{2}$H$^{+}$ and CCS and their ratio N(N$_{2}$H$^{+}$)/N(CCS) were plotted as a function of the evolved time (solid line with uncertainties, Figure \ref{fig:Model_density}). The discontinuities in the modeled column densities between different stages are due to the varying physical structures of the best-fit modelsfor each stage \citep{2014A&A...563A..97G,2015A&A...579A..80G}. 
For comparisons, the average values of measured results of N$_{2}$H$^{+}$, CCS, and their N(N$_{2}$H$^{+}$)/N(CCS) at different stages were also plotted in dotted line with uncertainties in Figure \ref{fig:Model_density}. It can be found that the modeled results show a similar trend with our measured results, i.e., increasing ratio of N(N$_{2}$H$^{+}$)/N(CCS) from HMSC, HMPO to UC H{\footnotesize II} stages.
This support the reliability of $N$(N$_{2}$H$^{+}$)/$N$(CCS) as a chemical clock of HMSFRs. 

\renewcommand\tabcolsep{30.0pt} 
\begin{deluxetable*}{lccc}
	\tablecaption{Derived parameters of the best-fit model of HMSC, HMPO and UC H{\footnotesize II}. \label{tab: best-fit model}}
	\tablehead{
		\colhead{Parameter} & 
		\colhead{HMSC} &
		\colhead{HMPO} &
		\colhead{UC H{\footnotesize II}} 
	}
	\startdata
	Density  & 7.78$\times$10$^5$ cm$^{-3}$ &4.50$\times$10$^5$ cm$^{-3}$ & 4.30$\times$10$^5$ cm$^{-3}$ \\
	Temperature  & 24.5 K & 103.7 K & 110.5 K \\
	Timescale & 19 179 yr & 32 990 yr& 34 286yr
	\enddata
\end{deluxetable*}

\section{Summary}\label{sec:summary}
In this paper, we presented the observations on molecular lines of N$_{2}$H$^{+}$, CCS $J_{N}$ = $8_{7}-7_{6}$ and $7_{7}-6_{6}$ toward 88 UC H{\footnotesize II}s with IRAM 30 m telescope, to check $N$(N$_{2}$H$^{+}$)/$N$(CCS) as a chemical evolutionary indicator of  HMSFRs. The main results in this work can be summarized as follows:

1.  
Among our 88 UC H{\footnotesize II}s, 87 and 33 sources were detected in the N$_{2}$H$^{+}$ $J$ = 1$-$0 and CCS $J_{N}$ = $8_{7}-7_{6}$ lines, respectively. All sources with detection of CCS $J_{N}$ = $8_{7}-7_{6}$ have detection of N$_{2}$H$^{+}$ $J$ = 1$-$0. Ten sources among those with detection of CCS $J_{N}$ = $8_{7}-7_{6}$ line were also detected in CCS $J_{N}$ = $7_{7}-6_{6}$ line.

2. For those sources with detections of CCS and N$_{2}$H$^{+}$ spectral lines, the line width of them was analyzed. It shows that the thermal broadening is not significant in both N$_{2}$H$^{+}$ and CCS lines. Comparisons show that the line width of CCS is normally larger than that of N$_{2}$H$^{+}$, suggesting that CCS is more likely from inner and more active star-forming regions.

3. For those sources with N$_{2}$H$^{+}$ $J$ = 1$-$0 detection, we estimated the optical depth of N$_{2}$H$^{+}$ $J$ = 1$-$0 using the line intensity ratio method and then obtained the excitation temperature and the column density of N$_{2}$H$^{+}$. 
Toward 10 sources with detections of two CCS lines, we determined the column density and the $T_{rot}$ of CCS, using the rotational diagram method. Using the average $T_{rot}$ of those 10 sources, we estimated the column density for other 23 sources with only CCS $J_{N}$ = $8_{7}-7_{6}$ detection. 

4. Through comparative analysis on measured results of HMSCs, HMPOs and UC H{\footnotesize II}s, we found that the column density ratio of N$_{2}$H$^{+}$/CCS increases from HMSCs to HMPOs, and then to UC H{\footnotesize II}s. This can be supported by our gas-grain chemical model, which shows the modeled column density ratio of N$_{2}$H$^{+}$ and CCS tends to increase with evolution age. And based on our best-fit model, we further constrained the mean physical properties and chemical age of HMFRS  to be 83 913 years, which is consistent with values from previously theoretical models on high-mass star formation.
Thus, we propose that $N$(N${_2}$H$^{+}$)/N(CCS) can be a reliable chemical evolutionary indicator in high-mass star formation regions.

\begin{acknowledgments}
	We thank the operators and staff at IRAM for their assistance during our observations. This work is supported by the Natural Science Foundation of China (No. 12373021, 12041302). J.X. G. thanks the Xinjiang Tianchi Talent Program (2024). Y. T. Y. and Y. X. W. are members of the International Max Planck Reisearch School (IMPRS) for Astronomy and Astrophysics at the Universities of Bonn and Cologne. H. Z. Y. would like to thank the China Scholarship Council (CSC) and the Ministry of Science and Higher Education of the Russian Federation (state contract FEUZ-2023-0019) for support. Based on data from the IRAM Science Data Archive. The partial HMSC data were obtained by Francesco Fontani with the IRAM 30-meter telescope under project 042-15.
\end{acknowledgments}

\bibliography{sample631}{}
\bibliographystyle{aasjournal}

\appendix
%
%

\section{Comparison of optical depth from two method}\label{sec: AppendixA}

\begin{figure*}[htbp]
	\centering
	{\includegraphics[width=8cm]{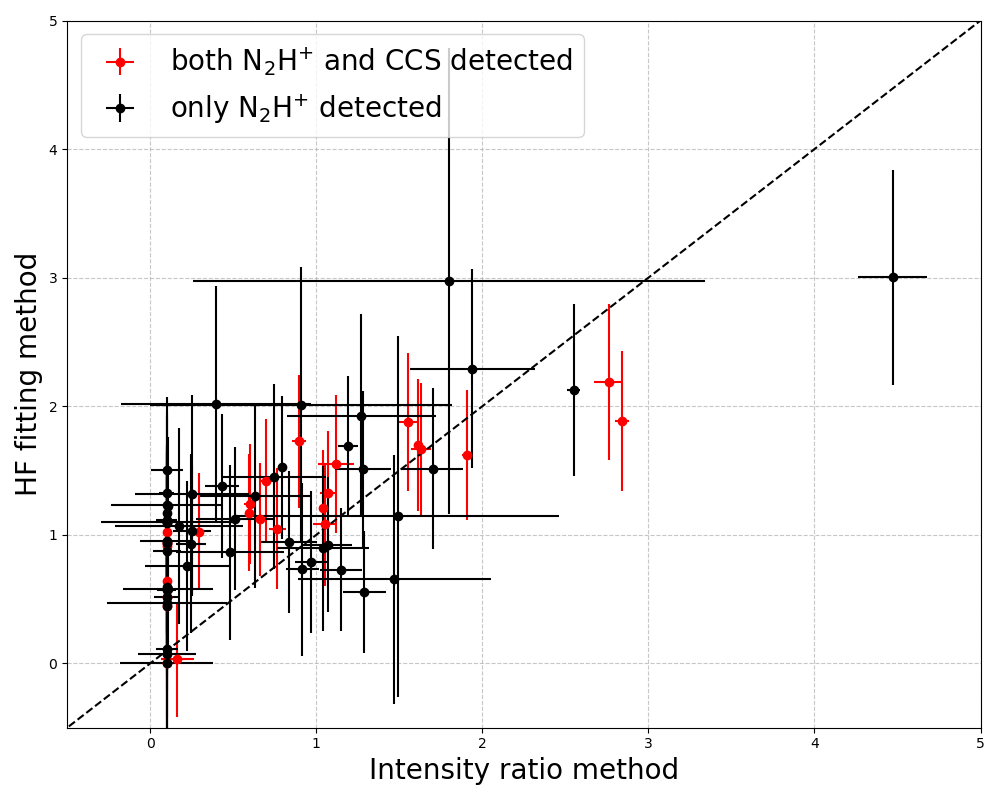}} 
	\caption{The comparison of the optical depth from the intensity ratio and HF fitting method. Sources with only N$_{2}$H$^{+}$ detection are in solid black circle, while sources with both N$_{2}$H$^{+}$ and CCS detections are in solid red circle. The black dashed line means that both optical depth have the same value.}
	\label{fig:optical_depth_comparison}
\end{figure*}

For the remaining 81 sources with N$_{2}$H$^{+}$ detection, we also estimated their optical depth by using the HF fitting method. A comparison of optical depth from two methods, i.e., the intensity ratio and HF fitting method, are showed in Figure \ref{fig:optical_depth_comparison}. This shows that the optical depth estimated from HF fitting method tends to be larger than that from intensity ratio method. However, for sources with both N$_{2}$H$^{+}$ and CCS detections (solid red circle), the mean optical depths from the intensity ratio and HF fitting method are 0.85 $\pm$ 0.05 and 1.16 $\pm$ 0.49, respectively, which are consistent with each other within uncertainties. 
Thus, we adopted the optical depths derived from the intensity ratio method in subsequent analyses, and accounted for the associated uncertainties through error propagation.

\section{The basic information of HMSC and HMPO data from Chen et al (2025 in prep.)}\label{sec: AppendixB}

Our data, the N$_{2}$H$^{+}$ ($J$ = 1-0) and CCS $J_{N}$ = $8_{7}-7_{6}$ and $7_{7}-6_{6}$ toward  HMSC and HMPO sample, was obtained from the ARO 12 m telescope observations. Observations were performed remotely from Guangzhou University, China, in 2024 March, April and May within project Chen\_24\_1 (PI: Jialiang Chen) and Chen\_24\_2 (PI: Jialiang Chen). The details of our data toward our  HMSC and HMPO sample are listed in Table \ref{tab:Chen_source list}. 

\startlongtable
\renewcommand\tabcolsep{10.0pt} 
		\begin{deluxetable*}{lccccc}
			\tablecaption{Basic Information of HMSC and HMPO Sample \label{tab:Chen_source list}}
			\tablehead{
				\colhead{Object}&
				\colhead{$\alpha$(2000)}&
				\colhead{$\delta$(2000)}&
				\colhead{D$_{sun}$}&
								\colhead{Evolutionary Stage}&
				\colhead{References}
				\\
				\colhead{}&
				\colhead{($^h \; ^m \; ^s$)}&
				\colhead{($^{\circ} \; ^{\prime} \; ^{\prime\prime}$)}&
				\colhead{(kpc)}&
								\colhead{}&
				\colhead{}
			}
			\decimalcolnumbers
			\startdata
			18182-1433-3  & 18:21:17.5 & -14:29:43 & 3.58  $\pm$ 0.54  & HMSC & {[}1{]} \\
			18223-1243-3  & 18:25:08.3 & -12:45:27 & 3.37  $\pm$ 0.51  & HMSC & {[}1{]} \\
			18247-1147-3  & 18:27:31.0 & -11:44:46 & 5.14  $\pm$ 0.77  & HMSC & {[}1{]} \\
			18337-0743-3  & 18:36:18.2 & -07:41:00 & 3.70  $\pm$ 0.56  & HMSC & {[}1{]} \\
			18337-0743-7  & 18:36:19.0 & -07:41:48 & 3.70  $\pm$ 0.56  & HMSC & {[}1{]} \\
			18385-0512-3  & 18:41:17.4 & -05:10:03 & 3.04  $\pm$ 0.46  & HMSC & {[}1{]} \\
			18530+0215-2  & 18:55:29.0 & +02:17:43 & 4.67  $\pm$ 0.70  & HMSC & {[}1{]} \\
			19175+1357-4e & 19:19:50.6 & +14:01:22 & 10.75  $\pm$ 1.61 & HMSC & {[}1{]} \\
			18151-1208    & 18:17:57.1 & -12:07:22 & 5.16  $\pm$ 0.77  & HMPO & {[}2{]} \\
			18223-1243    & 18:25:10.9 & -12:42:17 & 5.18  $\pm$ 0.78  & HMPO & {[}2{]} \\
			18264-1152    & 18:29:14.3 & -11:50:26 & 9.34  $\pm$ 1.40  & HMPO & {[}2{]} \\
			18308-0841    & 18:33:31.9 & -08:39:17 & 5.28  $\pm$ 0.79  & HMPO & {[}2{]} \\
			18454-0136    & 18:48:03.7 & -01:33:23 & 8.68  $\pm$ 1.30  & HMPO & {[}2{]} \\
			18488+0000    & 18:51:24.8 & +00:04:19 & 4.55  $\pm$ 0.68  & HMPO & {[}2{]} \\
			18530+0215    & 18:55:34.2 & +02:19:08 & 5.16  $\pm$ 0.77  & HMPO & {[}2{]} \\
			\enddata
			\tablecomments{
				Column(1): source name; column(2): R.A. (J2000); column(3): decl. (J2000); column(4): the heliocentric distance; column(5): evolutionary stage; column(6): references. [1] \cite{2005ApJ...634L..57S}, [2] \cite{2002ApJ...566..931S}.
			}
		\end{deluxetable*}

\section{Model column density calculation}\label{sec: Model column density calculation}

The abundance $[X]$ (with respect to H) of a species $X$ can be converted into column densities $N(X)$ with this method described in \cite{2016ApJ...830L...6J} : 
\begin{equation}\label{equ:X_N}
	\begin{aligned}
		N(X)&=2\times\sum_{i=2}^n\left(\frac{n(\text{H})_i[X]_i+n(\text{H})_{i-1}[X]_{i-1}}{2}\right)\\&\times(R_{i-1}-R_i),
	\end{aligned}
\end{equation}
where $R_i$ is the radius of the $i_{th}$ shell, and $R_1$ is the first shell as 0.5 pc from the center \citep{2014A&A...563A..97G}.  $n$ is the number of shells in the model ($n$ = 129).  $n(\text{H})_i$ is the gas density at radial point $i$ and $[X]_i$ the abundance of the species. Both $n(\text{H})_i$ and $[X]_i$ are constant within each model. 
The model column densities are averaged over the beam of the IRAM 30 m telescope ($\sim$27 \arcsec). The uncertainties on the modeled mean column densities were derived applying error propagation based on Equation (\ref{equ:X_N}). 



\end{document}